%% file: main.tex
\documentclass{aa}

\usepackage{graphicx}
\usepackage{txfonts}
\usepackage[acronym,shortcuts]{glossaries} 
\usepackage{booktabs}
\usepackage{bm}
\glsdisablehyper 
\usepackage{xspace}
\usepackage{hyperref}
\usepackage{orcidlink}
\usepackage{makecell}
\setcellgapes{3pt}
\newcommand{\Rsun}{\ensuremath{\,\rm{R}_{\odot}}\xspace}

\newcommand{\kms}{\ensuremath{\,\rm{km}\,\rm{s}^{-1}}\xspace}
\newcommand{\cms}{\ensuremath{\,\rm{cm}\,\rm{s}^{-1}}\xspace}
\newcommand{\Msun}{\ensuremath{\,\rm{M}_{\odot}}\xspace}

\newcommand{\AU}{\ensuremath{\,\mathrm{AU}}\xspace}

\newcommand{\yr}{\ensuremath{\,\mathrm{yr}}\xspace}

\newcommand{\erg}{\ensuremath{\,\mathrm{erg}}\xspace}
\newcommand{\ergs}{\ensuremath{\,\mathrm{erg}\,\rm{s}^{-1}}\xspace}
\newcommand{\ergg}{\ensuremath{\,\mathrm{erg}\,\rm{g}^{-1}}\xspace}
\newcommand{\cmsg}{\ensuremath{\,\mathrm{cm}^{2}\,\rm{g}^{-1}}\xspace}
\newcommand{\gcc}{\ensuremath{\,\mathrm{g}\,\rm{cm}^{-3}}\xspace}
\newcommand{\seconds}{\ensuremath{\,\mathrm{s}}\xspace}
\newcommand{\cm}{\ensuremath{\,\mathrm{cm}}\xspace}
\newcommand{\K}{\ensuremath{\,\mathrm{K}}\xspace}
\newcommand{\Phantom}{{\scshape phantom}\xspace }

\newcommand{\MESA}{{\scshape mesa}\xspace }

\newacronym{RSG}{RSG}{red supergiant}

\newacronym{CE}{CE}{common-envelope}
\newacronym[longplural={luminous red novae}, shortplural={LRNe}]{LRN}{LRN}{luminous red nova}

\newacronym{SPH}{SPH}{smoothed particle hydrodynamics}
\newacronym[longplural={equations of state}, plural={EoSs}]{EoS}{EoS}{equation of state}
\newacronym{LTE}{LTE}{local thermodynamic equilibrium}

\begin{document} 
\title{Common envelopes in massive stars}
\subtitle{III. The obstructive role of radiation transport in envelope ejection}

\author{Mike Y. M. Lau \inst{1}\fnmsep\thanks{\email{mike.lau@h-its.org}} \orcidlink{0000-0002-6592-2036} \and
        Ryosuke Hirai \inst{4,2,3} \orcidlink{0000-0002-8032-8174} \and
        Daniel J. Price \inst{2} \orcidlink{0000-0002-4716-4235} \and
        Ilya Mandel \inst{2,3} \orcidlink{0000-0002-6134-8946} \and
        Matthew R. Bate \inst{5} \orcidlink{0000-0002-2926-0493}
    }

\institute{
      Heidelberger Institut f\"{u}r Theoretische Studien, Schloss-Wolfsbrunnenweg 35, 69118 Heidelberg, Germany \and
      School of Physics and Astronomy, Monash University, Clayton, Victoria 3800, Australia \and
      OzGrav: The ARC Centre of Excellence for Gravitational Wave Discovery, Australia \and
      RIKEN Cluster for Pioneering Research (CPR), RIKEN, Wako, Saitama 351-0198, Japan \and
      Department of Physics and Astronomy, University of Exeter, Stocker Road, Exeter EX4 4QL, UK
    }


\abstract{
    We present 3D radiation hydrodynamics simulations of \ac{CE} evolution involving a 12\Msun red supergiant donor and a 3\Msun companion. Existing 3D simulations are predominantly adiabatic, focusing strongly on low-mass donors on the red giant and asymptotic giant branches. However, the adiabatic assumption breaks down once the perturbed \ac{CE} material becomes optically thin or when entering a longer-timescale evolutionary phase after the dynamical plunge-in. This is especially important for high-mass red supergiant donors, which have short thermal timescales, adding significant uncertainty to our understanding of how massive binary stars evolve into gravitational-wave sources, X-ray binaries, stripped-envelope supernovae, and more. We compare our radiation hydrodynamics simulations with an adiabatic simulation from Paper I that is otherwise identical, finding that radiative diffusion strongly inhibits \ac{CE} ejection. The fraction of ejected mass is roughly half that of the adiabatic case without accounting for recombination energy release. Almost no material is ejected during the dynamical plunge-in, and longer-timescale ejection during the slow spiral-in is suppressed. However, the orbital separation reached at the end of the dynamical plunge-in does not differ significantly. The large amount of remaining bound mass tentatively supports the emerging view that the dynamical plunge-in is followed by a non-adiabatic phase, during which a substantial fraction of the envelope is ejected and the binary orbit may continue to evolve.

    \glsresetall  
}


\titlerunning{Common envelopes in massive stars III}
\maketitle


\input{intro.tex}

\input{methods.tex}

\input{results.tex}

\input{discussion.tex}

\input{conclusion.tex}

\begin{acknowledgements}
	We thank Orsola De Marco, Miguel Gonz\'{a}lez-Bol\'{i}var, Luis Berm\'{u}dez-Bustamante, and members of the PSO and SET groups at the Heidelberg Institute for Theoretical Studies for useful discussions. M. Y. M. L. is supported by a Croucher Fellowship. Parts of this research were supported by the Australian Research Council Centre of Excellence for Gravitational Wave Discovery (OzGrav), through project number CE230100016. The simulations presented in this work were performed on the OzSTAR national facility at the Swinburne University of Technology. The OzSTAR program receives funding in part from the Astronomy National Collaborative Research Infrastructure Strategy (NCRIS) allocation provided by the Australian Government, and from the Victorian Higher Education State Investment Fund (VHESIF) provided by the Victorian Government.
\end{acknowledgements}

%
%
\bibliography{bibliography.bib}{}
\bibliographystyle{aa}

\begin{appendix}
	\input{resolution.tex}
    \input{test_problems.tex}
    \input{movies.tex}
\end{appendix}


\end{document}

%% file: intro.tex
\section{Introduction}
\label{sec:intro}
The long-standing basic picture of \ac{CE} evolution consists of a dynamical-timescale, adiabatic inspiral also known as the `plunge-in'\footnote{For a review of \ac{CE} evolution, see \cite{Ivanova+13,Ivanova+20}, \cite{Roepke+DeMarco23}, and \cite{Schneider+25}.}. This key assumption underlies the energy formalism \citep{van1976structure,Tutukov&Yungelson79,Webbink84} that equates the orbital energy released during the inspiral with the envelope binding energy. In line with this assumption, almost all 3D hydrodynamic simulations of the \ac{CE} plunge-in have neglected radiation hydrodynamics. Until recent years, these simulations have mainly targeted lower-mass donor stars on the red giant and asymptotic giant branches \citep[e.g.][]{Passy+2012,Ricker&Taam2012,Ohlmann+16,Chamandy+18,Reichardt+2019,Prust&Chang19,Sand+20,Gonzalez-Bolivar+22,Bermudez+24}. In such cases, radiative losses and thermal restructuring are assumed to be negligible because the donor envelope's thermal timescale is many orders of magnitude longer than the inspiral duration.

However, the assumption of adiabaticity is questionable for very massive \ac{RSG} donors, which have short thermal timescales \citep{Ricker+18,Vigna-Gomez+22}. More broadly, the evolution leading to the plunge-in may be non-adiabatic \citep{Pejcha+16b}, while the plunge-in itself may be followed by a slower, self-regulated inspiral phase \citep{Meyer+Meyer-Hofmeister79,Podsiadlowski01}, during which the luminosity from the orbital decay matches the radiated luminosity at the photosphere. These considerations call for incorporating radiation transport into 3D simulations, especially for massive donor stars.

However, this increases the computational cost of an already challenging calculation. Simulations of \ac{CE} evolution with \ac{RSG} donors have only become available in recent years \citep{Lau+22a,Lau+22b,Moreno+22,Vetter+24,Vetter+25,Landri+25}. All of these simulations are adiabatic, except for the preliminary results of a radiation hydrodynamics CE simulation by \cite{Ricker+18}, who modelled a 82.1\Msun \ac{RSG} donor and a 35\Msun companion black hole. More recently, \cite{Hatfull+24} presented hydrodynamic simulations with radiative emission and diffusion used to model the light curve of the V1309 Sco \ac{LRN} \citep{Tylenda+2011}, which is believed to be caused by the merger of a slightly evolved, $\approx 1.5\Msun$ star with a much lower-mass main-sequence companion. Other calculations incorporating energy transport were performed under 1D spherical symmetry \citep{Meyer+Meyer-Hofmeister79,Clayton+17,Fragos+19,Bronner+24}.

As a result, our understanding of the impact of radiation transport on \ac{CE} evolution, especially in \acp{CE} with massive-star donors, remains limited. Clarifying its role in the outcome of \ac{CE} interactions is essential for understanding the formation of compact binary systems. In massive binary stars, this includes gravitational-wave sources, X-ray binaries, and progenitors of stripped-envelope supernovae. Simulations with radiation hydrodynamics are also critical for bridging theoretical models of \acp{CE} with observations. Energy transport is needed to accurately model photospheric evolution, which is necessary for computing light curves that can be compared with observed \acp{LRN} \citep{Pastorello+19,Hatfull+24}. Radiation transport is also expected to influence the morphology and evolution of post-\ac{CE} bound ejecta and bipolar lobes \citep[e.g.][]{Vetter+24,Vetter+25,Gagnier&Pejcha23}, which likely contain optically thin material. Bipolar outflows in \acp{CE} are believed to contribute to the formation of planetary nebulae \citep[e.g.][]{DeMarco09,Garcia-Segua+18,Zou+20,Jacoby+21,Ondratschek+22}.

In this paper, we present 3D radiation hydrodynamics simulations of the \ac{CE} phase involving a 12\Msun \ac{RSG} donor and a 3\Msun companion. Radiation hydrodynamics is modelled using the flux-limited diffusion approximation. We compare these simulations with our previous adiabatic simulation, which used an otherwise identical setup, and find that radiation transport significantly obstructs envelope ejection, resulting in less than half the unbound mass compared to the adiabatic case. Notably, almost no material becomes unbound during the plunge-in phase itself and the sustained mass ejection previously seen during the slow spiral-in is strongly suppressed by radiative diffusion. However, we did not account for the release of recombination energy, which is known to significantly increase the amount of unbound envelope mass in adiabatic simulations.

In Sect. \ref{sec:methods}, we describe our implementation of radiative diffusion and the initial setup. We present our main results in Sect. \ref{sec:results}, showing the effect of radiation transport on ejecta morphology (\ref{subsec:ejecta}), the amount of unbound ejecta (\ref{subsec:unbound}), and the final orbital separation (\ref{subsec:final_sep}). To better understand these effects, in Sect. \ref{subsec:dynamical_effect}, we identify and analyse regions where radiation transport is dynamically important. In Sect. \ref{subsec:recombination}, we specifically examine the optical depth at the location of partially ionised hydrogen. In Sect. \ref{sec:discussion}, we discuss the limitations of our method (\ref{subsec:limitations}), compare our results with related studies (\ref{subsec:related_works}), and discuss implications of our findings for binary stellar evolution (\ref{subsec:implications}). Finally, we summarise our findings in Sect. \ref{sec:conclusion}.

%% file: methods.tex
\section{Methods}
\label{sec:methods}
\subsection{Overview}
\label{subsec:overview}
The initial binary system consists of a 12\Msun \ac{RSG} donor and a 3\Msun companion modelled as a point mass, which could represent either a main-sequence star or a black hole. As in our past simulations \citep[][hereinafter referred to as Papers I and II, respectively]{Lau+22a,Lau+22b}, we used the \ac{SPH} code \Phantom \citep{Price+18} with an implicit two-temperature flux-limited diffusion solver based on \cite{Whitehouse+Bate04} and \cite{Whitehouse+05} implemented for this work. Details of the method are explained in Sect. \ref{subsec:diffusion}. The initial hydrostatic 3D stellar model is identical to that in Paper I and is resolved with $2\times10^6$ \ac{SPH} particles. This enables a controlled comparison to infer the effects of radiation transport. We also performed an additional simulation in which the base of the \ac{RSG} envelope is heated to account for luminosity from the helium core and hydrogen-burning shell. This heating primarily offsets the slow increase in central density due to radiation transport and has an insignificant energy contribution over the course of the \ac{CE} phase (see Sect. \ref{subsec:nuclear}). 

As in Paper I, convection is initially suppressed in the \ac{RSG} envelope by using a stellar model with a flat entropy stratification. However, this setup does not prohibit convection from developing later, as shock heating and radiative diffusion modify the entropy structure during the \ac{CE} plunge-in. Convective flows in the bound ejecta have been observed in Papers I and II as well as in other studies \citep[e.g.][]{Ohlmann+2017,Vetter+24,Gagnier&Pejcha23}.

\subsection{Radiation hydrodynamics scheme}
\label{subsec:diffusion}
The implementation of implicit flux-limited diffusion in \Phantom was directly ported from the {\scshape sphng} code \citep{Benz+90,Bate+95,Bate+Keto15}. It follows the method developed by \cite{Whitehouse+Bate04} and \cite{Whitehouse+05}, implemented in {\scshape sphng} as described in \cite{Whitehouse+Bate06}. The radiation hydrodynamics equations we solve are given by \citep{Mihalas+Mihalas84,Turner+Stone01}
\begin{align}
    \frac{D\rho}{Dt} &= -\rho \nabla \cdot \mathbf{v}, \label{eq:cont} \\
    \rho \frac{D\mathbf{v}}{Dt} &= -\nabla p + \frac{\kappa\rho}{c} \mathbf{F}_\mathrm{rad} + \rho\bm{\Pi}_\mathrm{shock} - \rho\nabla\Phi, \label{eq:euler} \\
    \rho \frac{D\xi}{Dt} &= -\nabla\cdot\mathbf{F}_\mathrm{rad} - \nabla \mathbf{v}:\mathbf{P}_\mathrm{rad} + a_\mathrm{rad}c\kappa\rho(T_\mathrm{gas}^4-T_\mathrm{rad}^4), \label{eq:rad} \\
    \rho \frac{Du}{Dt} &= -p\nabla\cdot\mathbf{v} + \rho\Lambda_\mathrm{shock} - a_\mathrm{rad}c\kappa\rho(T_\mathrm{gas}^4-T_\mathrm{rad}^4), \label{eq:u}
\end{align}
where $D/Dt := \partial/\partial t + \mathbf{v}\cdot\nabla$ is the convective derivative and $\rho$, $\mathbf{v}$, $p$, $\mathbf{P}_\mathrm{rad}$, $\mathbf{F}_\mathrm{rad}$, $\kappa$, $\Phi$, $\xi$, $u$, $T_\mathrm{gas}$, and $T_\mathrm{rad}$ represent density, velocity, scalar isotropic pressure, radiation pressure tensor, radiation flux, opacity, gravitational potential, specific radiation energy, specific gas thermal energy, gas temperature, and radiation temperature, respectively. The radiation constant is $a_\mathrm{rad}$ and the speed of light is $c$. Equations (\ref{eq:euler})-(\ref{eq:u}) have been integrated over frequency, and we do not distinguish between the Planck mean, energy mean, and flux mean opacities, all of which are denoted by $\kappa$. The symbols $\bm{\Pi}_\mathrm{shock}$ and $\Lambda_\mathrm{shock}$ are shock dissipation terms \citep{Price12}. The latter usually includes a thermal conductivity term to correctly treat contact discontinuities \citep{Price08}. However, since we explicitly model physical energy conduction in the form of radiative diffusion, we disabled this term, so that $\Lambda_\mathrm{shock}$ only includes viscous shock heating. The final terms in Eqs. (\ref{eq:rad}) and (\ref{eq:u}) describe energy exchange via radiative absorption and emission. These terms use the expression for specific radiation energy, $\xi = a_\mathrm{rad}T_\mathrm{rad}^4/\rho$. We assume \ac{LTE}, meaning the gas is in thermal equilibrium with itself even if it is decoupled from the radiation field, allowing radiative emission to be written using the Planck function \citep[e.g.][]{Rybicki&Lightman86}. In the optically thick limit, there is complete thermodynamic equilibrium and $T_\mathrm{gas}=T_\mathrm{rad}$. In the optically thin limit, the gas and radiation can decouple, causing them to have different temperatures.

Equations (\ref{eq:cont})-(\ref{eq:u}) are closed by (i) an \ac{EoS}, (ii) an opacity law $\kappa(\rho,T_\mathrm{gas})$, (iii) an expression for the radiation flux, $\mathbf{F}_\mathrm{rad}$, and (iv) an expression for the radiation pressure tensor, $\mathbf{P}_\mathrm{rad}$. We adopt the \ac{EoS} for an ideal gas with a fixed adiabatic index of $\gamma=5/3$,
\begin{align}
    p = (\gamma-1)\rho u.
    \label{eq:eos}
\end{align}
The specific thermal energy is
\begin{align}
    u = \frac{\mathcal{R}T_\mathrm{gas}}{(\gamma-1)\mu},
\end{align}
where $\mathcal{R}$ is the gas constant and $\mu$ is the mean molecular weight, for which we adopt a constant value of $\mu = 0.61821$. We do not include the energy released by atomic and molecular recombination\footnote{In Paper II, we compared an adiabatic \ac{CE} simulation that used a fixed $\mu$ with one that varied $\mu$ according to the ionisation state, finding no significant differences in the evolution of orbital separation and the amount of unbound mass.} and the associated changes to $\mu$ and $\gamma$.

We used opacity tables from the \MESA stellar evolution code, as implemented in \Phantom by \cite{Reichardt+20}. These tables combine OPAL opacities for the high-temperature regime \citep{Rogers+96,Rogers+Nayfonov02} with SCVH opacities for the low-temperature regime \citep{Saumon+95}. They capture the opacity bumps associated with partially ionised hydrogen, helium, and iron, as well as molecular transitions occurring below $\approx 5000\K$. 

The radiative flux has the standard expression,
\begin{align}
    \mathbf{F}_\mathrm{rad} = -D\nabla E,
    \label{eq:Frad}
\end{align}
where $E=\rho\xi=a_\mathrm{rad}T_\mathrm{rad}^4$ is the volumetric radiation energy density. Radiation energy is transported (anti-)parallel to the temperature gradient $\nabla T_\mathrm{rad}$. The diffusion coefficient is given by $D = c\lambda/(\kappa\rho)$, where $\lambda$ is the flux limiter. We adopt the flux limiter from \cite{Levermore+Pomraning81}, which approaches $\lambda \rightarrow 1/3$ in the optically thick limit. In the optically thin limit, the flux limiter ensures that $|\mathbf{F}_\mathrm{rad}| \rightarrow cE$, so that radiation energy does not propagate faster than light.

The radiation pressure tensor is written as $\mathbf{P}_\mathrm{rad} = E \mathbf{f}$, where $\mathbf{f}$ is the Eddington tensor, whose components depend on the flux limiter $\lambda$. In the isotropic, optically thick limit, $\mathbf{P}_\mathrm{rad} \rightarrow E/3 \mathbf{I}$. In the optically thin limit, $\mathbf{P}_\mathrm{rad} \rightarrow E \hat{\bm{n}}\hat{\bm{n}}$, where $\hat{\bm{n}} = \nabla E/|\nabla E|$ is a unit vector aligned with the energy density gradient.

\subsection{Implicit solver for radiative diffusion}
\label{subsec:implicit}
Although \Phantom previously included an explicit solver for two-temperature radiative diffusion \citep{Biriukov+Price19}, an explicit scheme is only stable subject to a Courant condition on the maximum time step for flux diffusion, given by $\Delta t_\mathrm{rad} = C h^2\rho\kappa/(c\lambda)$ \citep{Whitehouse+Bate04}, where $h$ is the particle smoothing length and $C$ is a Courant number of order unity. It is a known issue that $\Delta t_\mathrm{rad}$ is often many orders of magnitude shorter than the Courant time step for hydrodynamic processes in optically thin regions, where $\Delta t_\mathrm{rad}$ approaches the light-crossing time of the resolution scale. These conditions are first encountered when the companion approaches the donor star's surface, raising tides and ejecting material in its wake. As a result, using an explicit solver for radiative diffusion in \ac{CE} evolution results in prohibitively small time steps, making long-term simulations computationally infeasible.

We address this problem by implementing an implicit solver for $u$ and $\xi$, based on \cite{Whitehouse+05}, that we directly ported from {\scshape sphng}. This solver remains stable even with arbitrarily large time steps. It uses the backwards Euler method to evolve the gas and radiation energy equations, solving them iteratively for each pair of \ac{SPH} neighbours using the Gauss-Seidel method until achieving the desired accuracy. Further details can be found in \cite{Whitehouse+05} and \cite{Bate+Keto15}. To avert a large number of random memory accesses, the implicit method requires building large arrays that cache interaction terms between \ac{SPH} neighbours. This significantly increases the memory requirement compared to the explicit scheme.

We set an accuracy tolerance of $10^{-6}$ for the solutions of $u$ and $\xi$, which we find to be a good compromise between accuracy and computational speed. Optically thick regions typically converge to this accuracy within a few iterations, whereas optically thin regions typically require considerably more iterations. As a result, computations become more expensive once optically thin ejecta are produced. We imposed a maximum of 250 iterations per time step and allowed the simulation to proceed even if the $u$ and $\xi$ solutions had not fully converged within the tolerance. To further improve computational efficiency, we optimised each iteration by only including \ac{SPH} particles that failed to converge in the previous iteration. This focuses most of the computational expense on particles in optically thin regions, which require the most iterations. We find that this optimisation does not lead to significantly different results in lower-resolution runs with 50,000 particles. In our simulations, most time steps converge within 10--20 iterations (see Sect. \ref{sec:results}). In Appendix \ref{app:tests}, we present the results of radiation hydrodynamics test problems from \cite{Whitehouse+05}.

\subsection{Initial setup and stellar model}
\label{subsec:initial}
The initial conditions of our simulations are identical to those of the `gas + radiation' \ac{EoS} simulation in Paper I. This was ensured by starting from the same simulation file. All material is therefore initially in thermodynamic equilibrium, with $T_\mathrm{gas}=T_\mathrm{rad}$. Below, we summarise key information about our 3D stellar model and refer the reader to Paper I for further details.

The donor star is a non-rotating 12\Msun \ac{RSG} with a radius of 618\Rsun. We do not drive equilibrium envelope convection in the pre-\ac{CE} \ac{RSG}, and so the envelope was constructed to have constant entropy for convective stability. The inner 3.84\Msun of the \ac{RSG}, comprising the helium core and radiative intershell, is replaced by a point mass that only interacts gravitationally with surrounding material. Its gravitational potential is softened using a cubic spline with a softening length of $h_1 = 9.25\Rsun$, transitioning to the exact Newtonian potential at $r_\mathrm{core} = 2h_1$. This boundary corresponds to the inner edge of the convective envelope in a \MESA stellar model evolved to core helium ignition. The 3\Msun companion, also modelled as a softened point mass, has a softening length of $h_2=2.15\Rsun$. The stars are initially placed in a circular orbit with a separation of 988\Rsun, where the \ac{RSG} radius exceeds the Roche radius by 25\%.

The 1D stellar profile was mapped to a 3D distribution of \ac{SPH} particles using stretch mapping and an asynchronous particle shifting technique that allows different regions of the star to relax on their local dynamical timescale \citep[Appendix A,][]{Lau+22a}. Paper I demonstrates that the 3D density structure is preserved to within $\lesssim 0.1\%$ when evolved in isolation for 90 times the surface free-fall time, even with just 50,000 \ac{SPH} particles (40 times fewer than in the production simulations).

\subsection{Thermal structure of the donor}
\label{subsec:thermal_structure}
\begin{figure}[h!]
    \includegraphics[width=\linewidth]{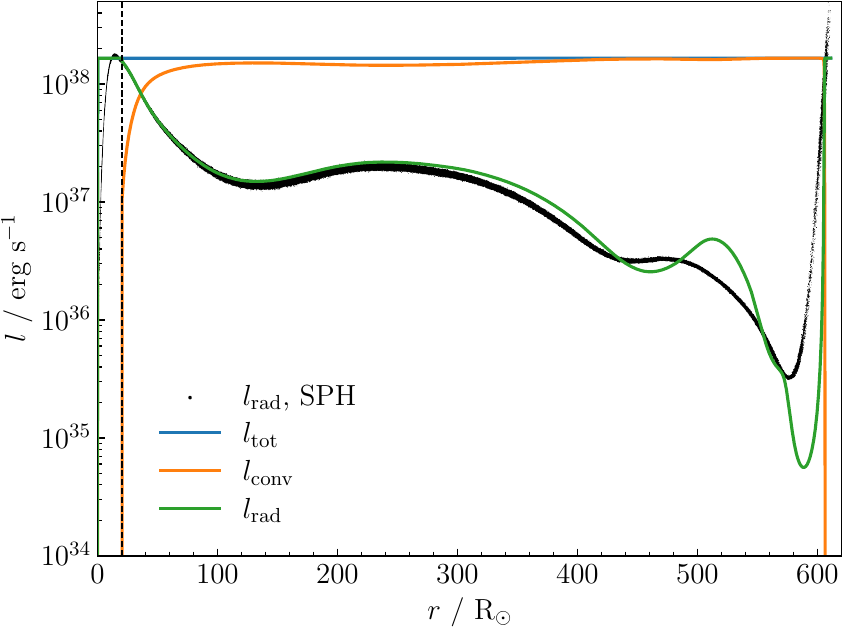}
    \centering 
    \caption{Comparison of the initial radiative luminosity profile ($l_\mathrm{rad}$) with that of a \MESA \ac{RSG} reference model. Each black dot shows one of two million \ac{SPH} particles. For the \MESA model, the blue line shows the total luminosity, the orange line shows the convective luminosity, and the green line shows the radiative luminosity. The dashed vertical line divides the convective envelope from the radiative interior.}
    \label{fig:lum}
\end{figure}

In a radiation hydrodynamics simulation, the donor star's initial thermal structure becomes an important consideration in addition to its mechanical structure. In thermal equilibrium, the total luminosity remains constant throughout the envelope of a giant star. Convection carries the luminosity generated through core and shell burning to the photosphere, where it is radiated away with the same luminosity. This is illustrated in Fig. \ref{fig:lum}, which shows the total, radiative, and convective luminosities of the reference \MESA \ac{RSG} model at the onset of core helium burning.

Our default simulation does not explicitly model cooling from the photosphere or heating from nuclear burning (though we conduct an additional simulation that includes nuclear luminosity; see Sect. \ref{subsec:nuclear}). To remain consistent with this approach, we initialised the envelope's entropy profile to be convectively stable, as described in Sect. \ref{subsec:initial}. At first order, the additional energy injected by the companion during the \ac{CE} inspiral must be transported via increased radiative or convective luminosity. We emphasise that our \ac{CE} simulations allow additional convection to develop self-consistently, as observed in Papers I and II.

With this approach, the start of the simulation only includes radiative diffusion as a means of energy transport. Figure \ref{fig:lum} shows the radiative luminosity of the 3D stellar model, $l_\mathrm{rad}(r) = 4\pi r^2 F_\mathrm{rad,r}$, where $F_\mathrm{rad,r}$ is the radial component of Eq. (\ref{eq:Frad}). Each black dot represents a single \ac{SPH} particle. The profile is qualitatively consistent with the reference \MESA model (shown in green), providing a reasonable representation of an \ac{RSG}'s radiative luminosity at the onset of core helium burning. The two profiles do not match exactly because the 3D model is based on the flat-entropy profile we constructed in Paper I, which only has the same total mass and radius as the \MESA model. However, this structure is not in thermal equilibrium because entropy flows into or out of regions where $\partial l_\mathrm{rad}/\partial r$ is negative or positive, respectively. Specifically, there are three dips in the radiative luminosity profile that coincide with opacity bumps from partially ionised iron, HeII, and HeI (ordered from smaller to larger radii) in the OPAL opacity tables. Also, below $r_\mathrm{core} = 18.5\Rsun$ (marked by the dashed vertical line), the softened density profile results in a much smaller radiative luminosity compared to the \MESA model, which maintains a flat $l_\mathrm{rad}$ profile down to the hydrogen-burning shell. Consequently, the star is expected to undergo thermal adjustments at these opacity bumps and below $r_\mathrm{core}$ on the local radiative diffusion timescale.

Indeed, we find that when evolving the 3D \ac{RSG} without a companion, entropy accumulates at the base of the convective envelope. This is associated with a gradual increase in central density. At $t=4.2\yr$, corresponding to 15 times the surface free-fall time, a buoyant plume of material begins to rise, marking the onset of convection despite an initially flat entropy profile. By this point, the central density has increased by $25\%$. However, this rising plume is not expected to introduce major artefacts to our initial conditions, since by 4.2\yr, the companion in our \ac{CE} simulation has already plunged deep into the \ac{RSG} envelope (see Sect. \ref{sec:results}). Also, the plume has a minor impact on the envelope density and pressure, as convection is highly efficient in most of the envelope.

\subsection{Nuclear burning luminosity}
\label{subsec:nuclear}
To assess the potential impact of the slow rise in central density, we conducted an additional simulation where we heated the base of the \ac{RSG} envelope to compensate for the energy transported outwards. In a real \ac{RSG}, this luminosity is generated by core helium burning and hydrogen shell burning. We applied a luminosity of $L_\star=1.7\times 10^{38}\ergs$, matching that of the \MESA model, to gas particles within a radius $r_\mathrm{core}$ of the point mass representing the donor's core. This heating was applied uniformly in mass, meaning each \ac{SPH} particle's thermal energy\footnote{The injected energy is distributed between the gas and radiation fields as they are in thermodynamic equilibrium deep within the star.} increased at the rate,
\begin{align}
    \frac{du_i}{dt} = \frac{L_\star}{m(r_\mathrm{core})},
\end{align}
where $m(r_\mathrm{core})$ is the total gas mass interior to $r_\mathrm{core}$. We find that the central density of the heated \ac{RSG} only increases by 6\% over the same 4.2\yr duration. The total energy injected by this luminosity over the course of the \ac{CE} inspiral is only a few percent of the envelope's initial gravitational binding energy, $E_\mathrm{bind}$,
\begin{align}
    \frac{L_\star \Delta t_\mathrm{inspiral}}{E_\mathrm{bind}} = 0.022 \biggl( \frac{L_\star}{1.7\times10^{38}\ergs} \biggr) \biggl(\frac{\Delta t_\mathrm{inspiral}}{5\yr}\biggr) \label{eq:lnuc} \\
    \biggl( \frac{E_\mathrm{bind}}{1.2\times10^{48}\erg}\biggr)^{-1}. \nonumber
\end{align}
Here, we assumed a duration of $\Delta t_\mathrm{inspiral} = 5\yr$, roughly the time at which the rapid plunge-in terminates (see Sect. \ref{sec:results}). As such, we do not expect significant qualitative differences between the outcomes of the heated and non-heated simulations. This is consistent with \cite{Fragos+19}, who performed a 1D \ac{CE} simulation with a 12\Msun \ac{RSG} and found that the cumulative nuclear energy contributes only a few percent of the change in gravitational potential energy. The `heated' simulation primarily serves as a control, where the central density and entropy of the envelope are better maintained. We do not consider the heated simulation to be our default calculation, as our setup also suppresses the equilibrium convection that would normally transport the nuclear luminosity, resulting in an inconsistency. For future \ac{CE} simulations, it would be valuable to explore the effects of a more realistic setup where initial envelope convection is driven via central heating and cooling from a resolved photosphere.

%% file: results.tex
\section{Results}
\label{sec:results}
We report the results of three different \ac{CE} simulations: (i) radiative (radiative diffusion without central heating), (ii) radiative + heating (radiative diffusion with central heating), and (iii) adiabatic (the `gas + radiation' \ac{EoS} simulation in Paper I). The first two cases are also collectively referred to as the radiative simulations. Below, we provide an overview of the simulations that is common to all three cases.

\begin{figure*}
    \centering
    \includegraphics[width=\linewidth]{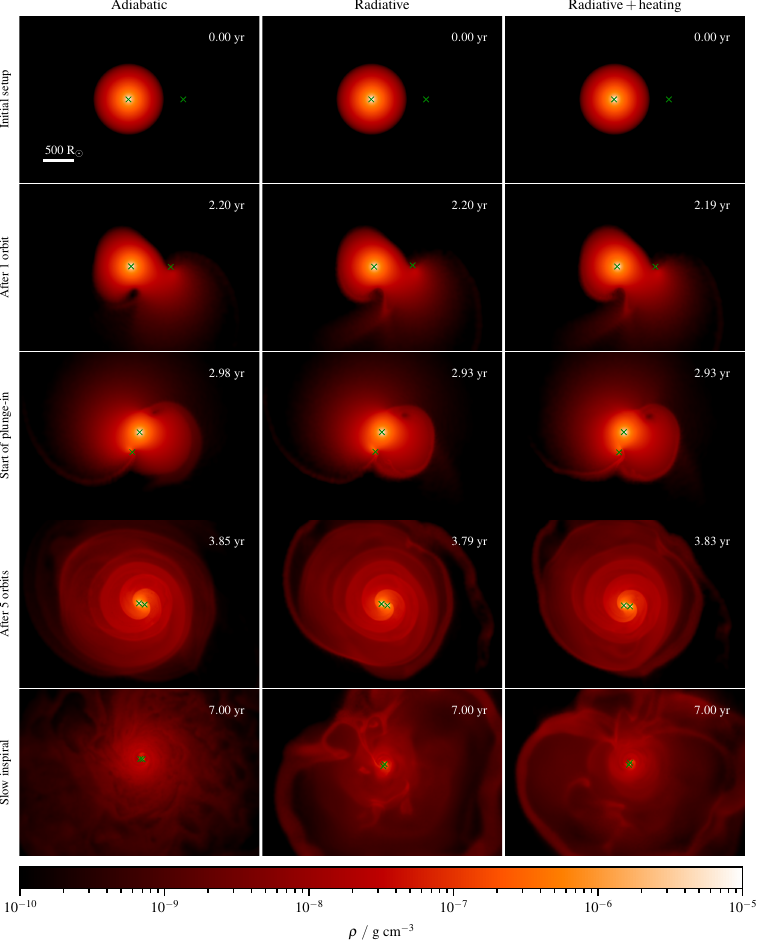}
    \caption{Density in the initial orbital plane ($z=0$) in 4320-by-3000\Rsun panels. The crosses mark the locations of the donor's core and the companion. Each column corresponds to a different simulation while each row corresponds to a different stage in the evolution. Row 5 shows the slow inspiral taking place several years after the plunge-in. The inspiral timescale, $t_\text{inspiral} = a/(-\dot{a}$), is several hundred to a thousand times longer than the orbital period. This figure was made with the \textsc{Sarracen} package \citep{Harris+Tricco23}. Online movies of our simulations are listed in Table \ref{tab:movies}.}
    \label{fig:rho_xy}
\end{figure*}

\begin{figure*}
    \centering
    \includegraphics[width=\linewidth]{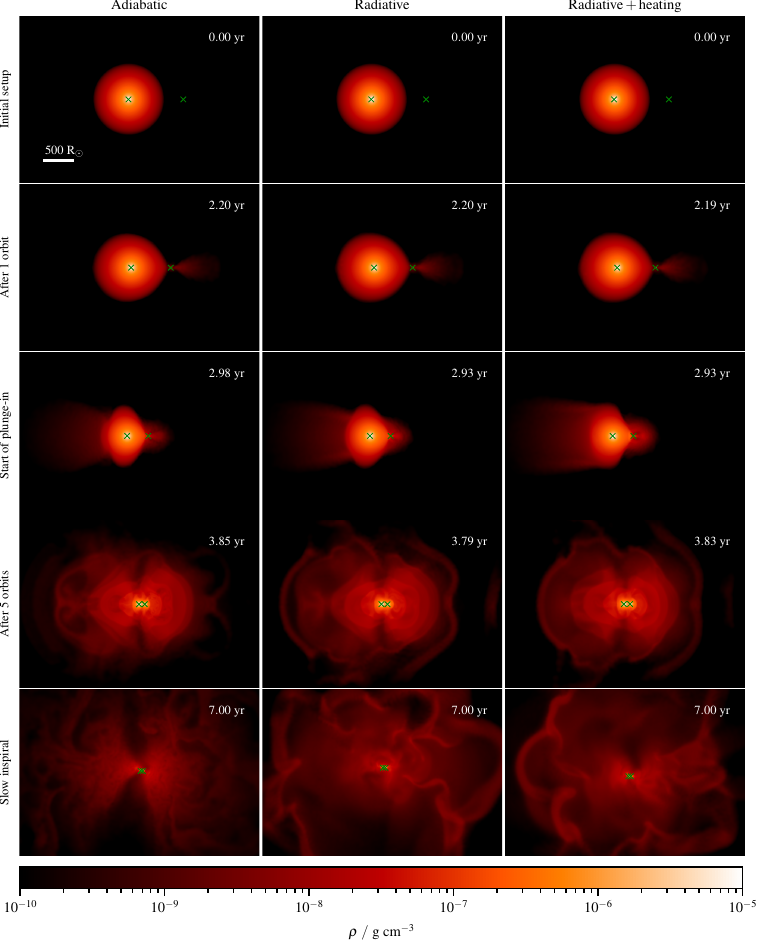}
    \caption{Same as Fig. \ref{fig:rho_xy}, except each 4320-by-3000\Rsun panel shows a slice viewed edge-on that contains the stellar cores. Online movies of our simulations are listed in Table \ref{tab:movies}.}
    \label{fig:rho_xz}
\end{figure*}

Figures \ref{fig:rho_xy} and \ref{fig:rho_xz} display density slices from the three simulations at different stages of the \ac{CE} inspiral. Row 1 shows the initial setup, where the orbital separation is 988\Rsun and the \ac{RSG} overfills its Roche lobe by 25\%. Initially, the implicit radiation solver converges to the specified tolerance within a few to ten iterations. After one orbit (row 2), the companion tidally deforms the \ac{RSG} and ejects material in its wake. As optically thin ejecta are produced, the solver requires around 30 iterations to converge. The companion enters the \ac{RSG} envelope at around 3\yr (row 3) and spirals inwards rapidly due to gravitational drag. Row 4 shows the spiral shocks generated by the motion of the two stellar cores in the \ac{CE}, causing some of the envelope material to expand and eventually be ejected. The orbital separation stabilises as the surrounding envelope material drops in density and reaches a high degree of co-rotation with the stellar cores. The inspiral timescale may be defined as
\begin{align}
    t_\mathrm{inspiral} = \frac{a}{-\dot{a}},
\end{align}
where $a$ is the separation between the donor's core and the companion. This timescale becomes several orders of magnitude longer than the orbital period ($P_\mathrm{orb}/t_\mathrm{inspiral} \lesssim 10^{-3}-10^{-2}$). Most of the envelope remains bound at this stage, forming an extended, diffuse structure (row 5). During this phase, some time steps required an even larger number of iterations to reach convergence, and $\sim 0.1\%$ of time steps did not converge after 250 iterations, the maximum number allowed. This represents a very small fraction of time steps and likely only reflects non-convergence in distant, optically thin particles, where the diffusion approximation breaks down in any case. During the plunge-in, a small number of distant, unbound \ac{SPH} particles (at least 30,000\Rsun away from the centre of mass) were removed because these low-density ($\rho\lesssim 10^{-15}\gcc$) particles had smoothing kernels that overlapped with a significant fraction of all particles in the simulation, causing an unbalanced load in the density calculation. This removal resulted in negligible mass loss, less than $4\times10^{-3}\%$ of the total.

Energy non-conservation\footnote{This excludes energy explicitly injected into the heated simulation.} exceeded 10\% near $t=8\yr$ for the radiative simulations (see Appendix \ref{app:conservation}), and the simulations were terminated shortly after. In contrast, the adiabatic simulation conserved total energy to within 0.05\% up to $t=11.6\yr$. The much larger energy accumulation in the radiative cases occurs because the implicit method does not conserve energy exactly.

\subsection{Ejecta morphology}
\label{subsec:ejecta}

\begin{figure*}
    \centering
    \includegraphics[width=\linewidth]{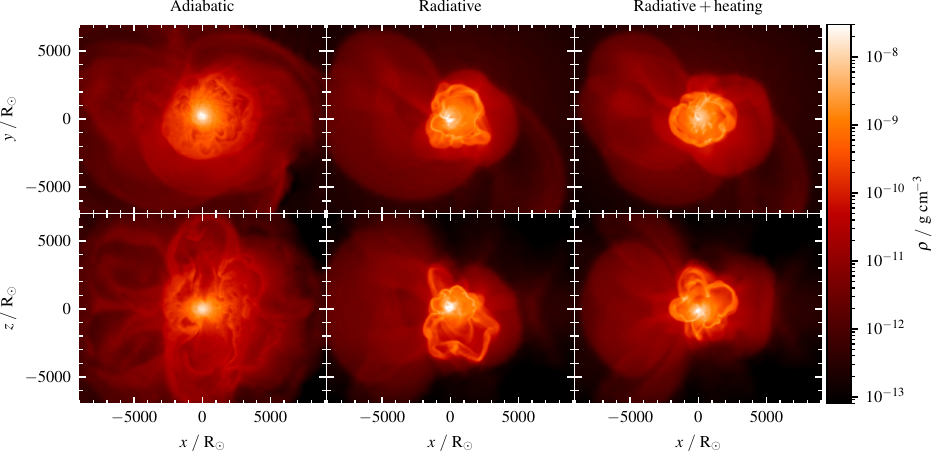}
    \caption{Same as Figs. \ref{fig:rho_xy} and \ref{fig:rho_xz}, but showing the late-time ejecta ($t=7.71\yr$) with a larger scale (18,200-by-14,000\Rsun) and a lower-density colour map. Top row: Orbital-plane slices. Bottom row: Edge-on slices.}
    \label{fig:ejecta}
\end{figure*}

The addition of radiation transport leads to the formation of more condensed density structures during and after the plunge-in. For instance, rows 4 and 5 of Figs. \ref{fig:rho_xy} and \ref{fig:rho_xz} show that the outer regions of the \ac{CE} form dense outer shells in the radiative simulations. In the adiabatic simulation, thermal energy deposited by the bow shock is fully utilised to expand the downstream envelope gas. However, in the non-adiabatic cases, radiative diffusion can remove thermal pressure, reducing the work done by the gas and even causing compression.

The last rows of Figs. \ref{fig:rho_xy} and \ref{fig:rho_xz} show the ejecta morphology after the plunge-in. In all three simulations, the equatorial region is primarily occupied by bound turbulent gas, while the regions above and below the orbital plane are more evacuated. In the adiabatic case, the bound equatorial material exhibits finer and more strongly stratified flows. This is consistent with the expectation that any energy conduction or diffusion (in this case, radiative diffusion) smooths out density and temperature contrasts. Only larger-scale turbulent flows, which are less strongly stratified, are preserved in the radiative simulations\footnote{While additional thermal conductivity is normally used for treating contact discontinuities in \ac{SPH}, this has not been used in our setup, as stated in Sect. \ref{subsec:diffusion}.}.

In fact, in the radiative simulations, the dense outer shells of the inflated \ac{CE} are susceptible to the Rayleigh-Taylor instability. We observe the formation of inward-extending fingers that later evolve into the characteristic mushroom-shaped downflows near the end of the plunge-in, continuing to the end of the simulation. However, these flow patterns are disrupted by outwardly moving spiral shocks and become mixed with the bound turbulent gas. This is illustrated in the last rows of Figs. \ref{fig:rho_xy} and \ref{fig:rho_xz}\footnote{This can be seen more clearly in videos of the simulations.}.

To show the full extent of the \ac{CE} ejecta near the end of the simulations, we present Fig. \ref{fig:ejecta}, showing face-on and edge-on density slices on a scale that is more than four times larger and with a colour map that displays lower densities. In the adiabatic case, low-density bipolar outflows are launched at $t\approx 6\yr$ and expand homologously at speeds of $\approx 200\kms$. These outflows are hydrodynamically collimated by the equatorially concentrated material that is still bound, inflating bipolar lobes (a phenomenon also described in Paper II). They lead to a persistent increase in the amount of unbound envelope mass following the plunge-in. Bipolar outflows in \ac{CE} evolution have also been recently discussed by \cite{Vetter+24,Vetter+25}, who performed magnetohydrodynamic \ac{CE} simulations resulting in faster and more strongly collimated jets. In their case, the jetted expulsion only accounts for $\sim 10\%$ of the total ejected envelope mass. They similarly report bound material that is equatorially concentrated in a centrifugally supported structure.

\begin{figure*}
    \centering
    \includegraphics[width=\linewidth]{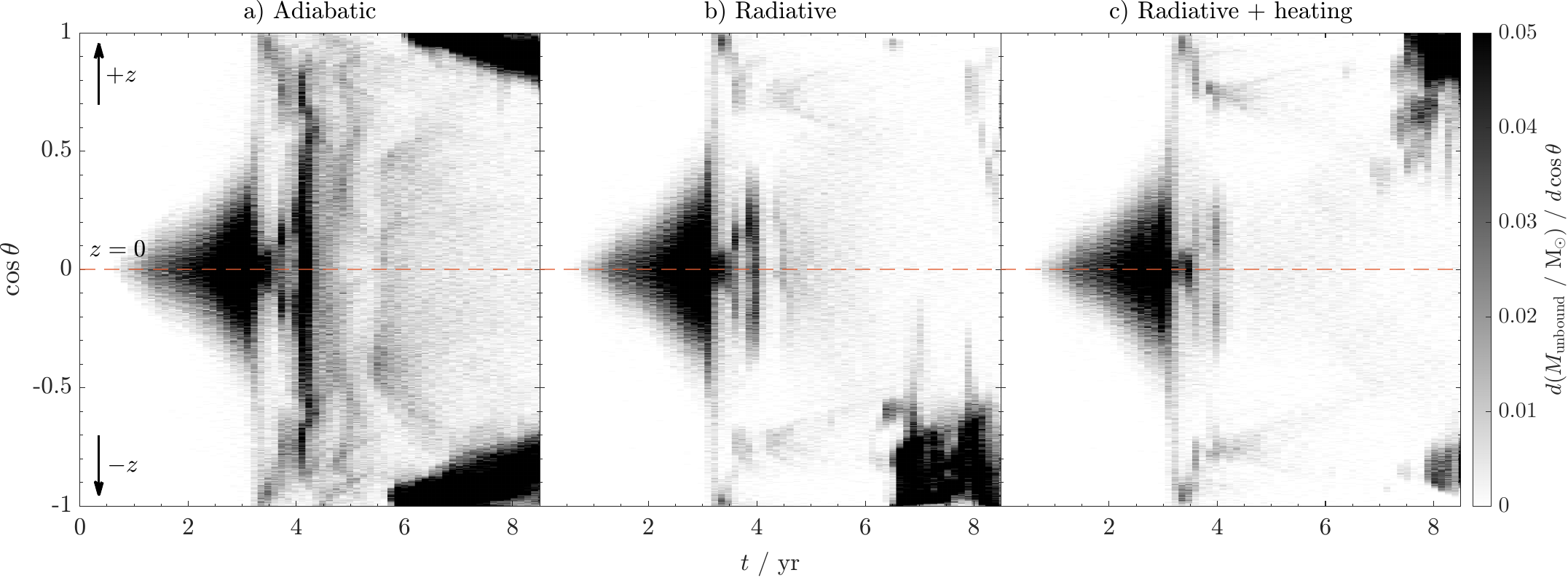}
    \caption{Angular distribution of particles that become unbound between consecutive simulation snapshots, meaning the sums of their kinetic and potential energies become positive. The angle $\theta$ is the polar angle measured from $\hat{\mathbf{z}}$, which is the direction of the initial orbital angular momentum vector. Material is initially ejected near the orbital plane, then closer to the poles in single or double-sided outflows after the plunge-in.}
    \label{fig:unbound_theta}
\end{figure*}

In the `radiative' and `radiative + heating' simulations, poloidal outflows are also observed a few years after the plunge-in. However, these outflows are weaker and less distinctly separated from the equatorial ejecta. Figure \ref{fig:unbound_theta} shows the angular distribution of newly unbound ejecta in all three simulations as a function of time. In the `radiative' simulation, two downward plumes of energetically unbound material are ejected from the central binary starting at $t\approx 6.4\yr$. The single-sided and episodic nature of this ejection reflects obstruction of the outflow by surrounding envelope gas that has fallen back. Similar outflows were seen in the simulations of the post-dynamical inspiral evolution by \cite{Gagnier+Pejcha25}. As we show in Sect. \ref{subsec:unbound}, the radiative simulations contain more fallback material because less of the \ac{CE} is ejected. Furthermore, radiation transport is able to concentrate this material by removing pressure support. In the `radiative + heating' simulation, an upward plume and a weaker downward plume emerge instead at a slightly later time of $t\approx 7.4\yr$. These differences are not expected to be systematic effects of central heating, as the exact launch time and direction of a weak outflow are sensitive to the surrounding turbulent ejecta, particularly the equatorial material that primarily contributes to collimation.

\subsection{Unbound mass}
\label{subsec:unbound}
\begin{figure}
    \centering
    \includegraphics[width=\linewidth]{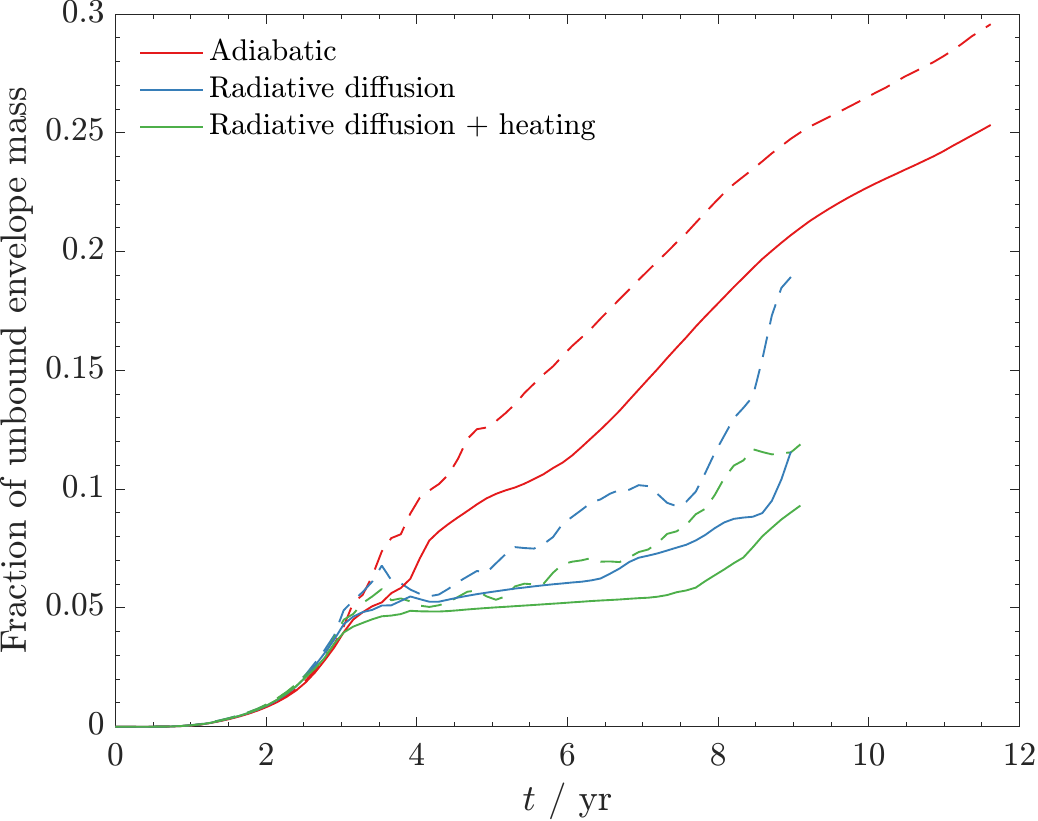}
    \caption{Fraction of unbound envelope mass as a function of time for the adiabatic simulation (red), the `radiative' simulation (blue), and the `radiative + heating' simulation (green). Solid lines assume material is unbound when the total mechanical energy is positive ($e_\mathrm{k} + e_\mathrm{p} > 0$), while dashed lines are less restrictive, considering material to be unbound when the sum of the mechanical and internal energy is positive ($e_\mathrm{k} + e_\mathrm{p} + e_\mathrm{gas} + e_\mathrm{rad}  > 0$).}
    \label{fig:unbound}
\end{figure}

\begin{figure}
    \centering
    \includegraphics[width=\linewidth]{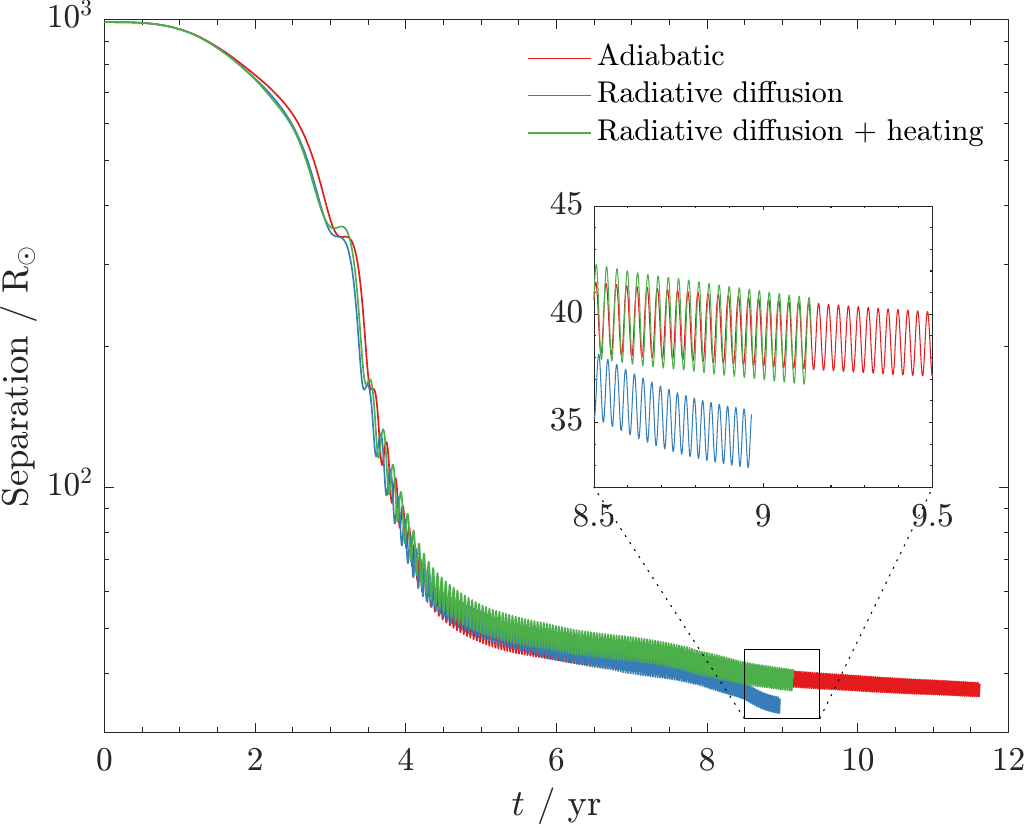}
    \caption{Separation between the \ac{RSG} core and the companion as a function of time for the adiabatic simulation (red), the `radiative' simulation (blue), and the `radiative + heating' simulation (green).}
    \label{fig:sep}
\end{figure}

Figure \ref{fig:unbound} shows the evolution in the fraction of unbound envelope mass. Material is considered unbound if the sum of its kinetic and potential energy is positive. For comparison, we also plot results assuming an alternative criterion that includes gas and radiation energy (dashed lines).

As with past 3D simulations of \ac{CE} evolution that do not include recombination energy, only a small fraction of the total envelope is unbound within a few years to a decade after the plunge-in, which is often the end of such simulations. At $t=9\yr$, the adiabatic simulation records only 20.8\% (1.7\Msun) of the \ac{RSG} envelope as unbound, increasing at a rate of $0.20~\mathrm{M}_\odot~\mathrm{yr}^{-1}$ (see Paper I). Including radiative diffusion reduces the amount of unbound ejecta further. By a similar time, only 11.0\% (0.90\Msun) of the envelope is ejected in the `radiative' simulation, while 8.9\% (0.72\Msun) of the envelope is ejected in the `radiative + heating' simulation. In both radiative cases, the amount of unbound mass is less than half of that in the adiabatic case. However, note that the radiative simulations display ongoing mass ejection at late times that is linked to the intermittent plumes discussed in Sect. \ref{subsec:ejecta}.

In the adiabatic simulation, we identify three phases of mass ejection in Fig. \ref{fig:unbound}. The first phase occurs prior to the plunge-in ($t\lesssim 3\yr$, during which material is ejected in the companion's wake. In this phase, the evolution in the fraction of unbound envelope mass is nearly identical across all three simulations. In each case, 4--5\% of the envelope mass (0.32--0.41\Msun) becomes unbound, with the ejection rate rising as the companion approaches the \ac{RSG} surface. This growth is abruptly halted when the companion enters the distorted \ac{RSG} envelope, cutting off the outflow of high-velocity ejecta through the $L_2$ point.

A second ejection event occurs at $t\approx 4\yr$ in the adiabatic simulation, associated with energy injected by spiral shocks during the \ac{CE} plunge-in. As shown in Fig. \ref{fig:unbound_theta}a, the ejecta span a much larger angular range due to the \ac{CE} homogenising energy deposited deep inside. However, the total amount of unbound ejecta remains a small fraction ($<10\%$) of the total envelope mass, consistent with previous adiabatic \ac{CE} simulations\footnote{Adiabatic simulations that include recombination energy release suggest that a large fraction of the envelope may be ejected \citep[e.g.][]{Nandez+15,Sand+20,Lau+22a,Lau+22b,Moreno+22}, but only if the simulations extend well beyond the dynamical inspiral phase.}. This finding has long contrasted with the widely adopted picture of dynamical \ac{CE} ejection, and the radiative simulations exacerbate this disagreement because this second ejection event is almost entirely absent. Instead, the ejection rate remains small (a few times $10^{-2}~\mathrm{M}_\odot~\mathrm{yr}^{-1}$) for several years after the plunge-in until intermittent outflows set in. Shortly after the plunge-in begins, the total unbound envelope mass in the radiative simulation is $\approx 1\%$ higher in absolute fraction than in the centrally heated simulation. As shown in Eq. (\ref{eq:lnuc}), this variation is unlikely to be a direct effect of energy injection. Instead, it likely results from the difference in the \ac{RSG}'s ability to maintain its central density depending on whether or not luminosity is injected in the centre, as discussed in Sects. \ref{subsec:thermal_structure} and \ref{subsec:nuclear}.

In the adiabatic simulation, a third phase of mass ejection proceeds from plunge-in, beginning around $t\approx 6\yr$. Material is ejected through the bipolar outflows described in Sect. \ref{subsec:ejecta}. A similar process occurs in the simulations with radiation transport, but a clear bipolar morphology is absent. Instead, ejection happens through single-sided and intermittent plumes. The onset of this ejection is delayed by 2--3 years compared to the adiabatic case, resulting in a much lower total fraction of unbound envelope mass at the same time. Due to energy non-conservation at late times, we have not followed the simulations long enough to determine whether this form of envelope ejection can be sustained. The episodic nature of these outflows is reflected in the uneven rise in the amount of unbound mass, particularly in the `radiative' case (blue curve). In the radiative simulations, the amount of unbound ejecta computed including thermal and radiation energies (dashed lines) varies irregularly with time. Dips appear at certain points, which indicate instances where ejected material has lost energy upon becoming optically thin and is subsequently considered energetically bound again.

\subsection{Final separation}
\label{subsec:final_sep}
Figure \ref{fig:sep} shows the evolution of the separation between the \ac{RSG} core and the companion. Despite differences in the amount of ejected mass, radiative diffusion has little impact on the orbital separation up to the end of the plunge-in phase. The orbital separations remain within $\approx 10\%$ of each other until $t=4\yr$, even though noticeable differences in the amount of unbound mass emerge around $t\approx 3.8\yr$. By $t=8\yr$, near the end of the radiative simulations, the orbital semi-major axes in the adiabatic, `radiative', and `radiative + heating' simulations are 40.9, 39.4, and 42.1\Rsun, respectively. These differences are minor and are comparable to the orbital eccentricities of $\approx 0.05$ recorded at the same time. In contrast, our previous works found that variations in envelope thermal energy (Paper I) and different sources of recombination energy (Paper II) resulted in more discernable changes to the semi-major axis near the end of the plunge-in. 

At the end of the simulations, the semi-major axis continues to decrease at a rate of a few times 0.01--0.1\% per orbit. In Papers I \& II, we recorded the post-plunge-in orbital separation when this rate dropped to 0.05\% per orbit ($P_\mathrm{orb}/t_\mathrm{inspiral} < 5\times10^{-4}$). However, in the radiative simulations, the rate of spiral-in does not slow steadily, making it unclear whether the orbital semi-major axis is approaching a stable value. In particular, the `radiative' simulation shows a pronounced decrease in orbital separation around $t\approx 8.5\yr$, coinciding with a relatively steep increase in unbound mass. The cause of this decrease is unclear and is plagued by numerical difficulties associated with simulating the late stages of the slow spiral-in, including a lack of spatial resolution, the softening length of the point masses becoming a large fraction of their Roche radii, and a potentially problematic buildup of (non-conserved) total energy, reaching $\approx 10\%$ at this point. Given these issues, it would be more instructive to study this behaviour in simulations targeting the post-plunge-in evolution \citep[e.g.][]{Gagnier+Pejcha25}, if it indeed occurs.

Despite these uncertainties, it is still possible to conclude that the orbital separation during and shortly after the plunge-in is not significantly affected by the inclusion of radiative diffusion. This is likely because the final separation and orbital evolution inside the \ac{CE} are primarily influenced by deeper, optically thick layers of the envelope, where radiative diffusion is not dynamically important. This aligns with analogous behaviour studied in Paper II, where we found that energy released at large radii from hydrogen recombination did not affect the post-plunge-in orbital separation despite contributing to the ejection of substantially more mass, whereas helium recombination energy, released much deeper, increased the final separation by $\approx 16\%$.

\subsection{Dynamical effect of radiation transport}
\label{subsec:dynamical_effect}
\begin{figure*}
    \centering
    \includegraphics[width=\linewidth]{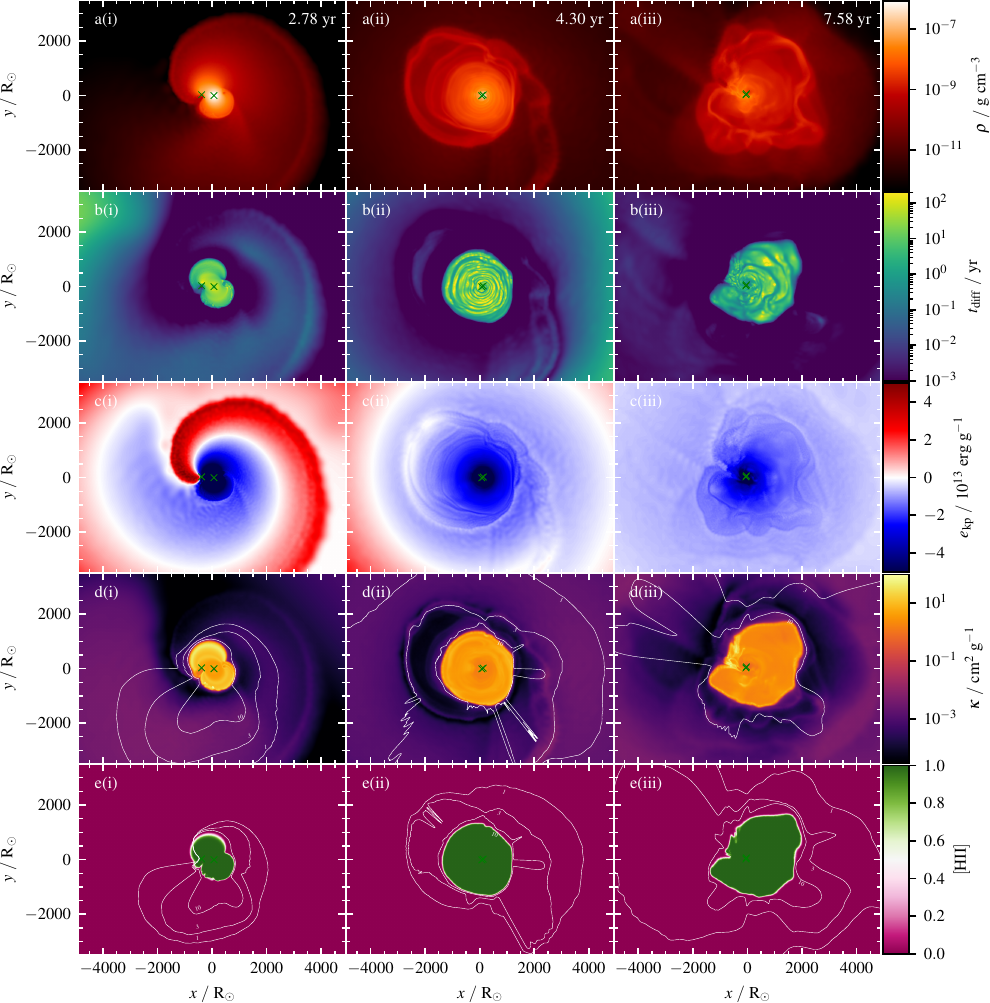}
    \caption{Top to bottom: Various quantities in orbital-plane slices from the `radiative' simulation: density ($\rho$), local diffusion time ($t_\mathrm{diff}$, Eq. (\ref{eq:tdiff})), kinetic plus potential energy per unit mass ($e_\mathrm{kp}$), opacity ($\kappa$), and the density-weighted hydrogen ionisation fraction ([HII]). Left to right: Snapshots representative of the pre-plunge-in phase, the end of the rapid plunge-in, and the late-time slow spiral-in. In rows (d) and (e), we overlay contours where the optical depth equals 1, 3, and 10, from largest to smallest size. The contours in row (d) include the contribution from all material, whereas those in row (e) include only the contribution from bound material ($e_\mathrm{kp} > 0$). The fact that these contours enclose the hydrogen partial ionisation zone (marked by the sharp transition between green and pink) suggests that a significant fraction of hydrogen recombination energy could be absorbed by the bound material.}
    \label{fig:tdiff}
\end{figure*}

To better understand the differences between the adiabatic and radiative simulations, we identify regions within the \ac{CE} and ejecta where radiation transport is efficient. Figure \ref{fig:tdiff} shows various quantities at three representative snapshots in the `radiative' simulation, including density ($\rho$), a local diffusion timescale ($t_\mathrm{diff}$), specific energy ($e_\mathrm{kp}$, the sum of kinetic and potential energy), opacity ($\kappa$), and hydrogen ionisation fraction ([HII]). The $e_\mathrm{kp}$ plots depict bound material in blue, unbound material in red, and the transition region in white. Similarly, the [HII] plots show ionised hydrogen as green, neutral hydrogen as pink, and the narrow partial ionisation zone as white. The local diffusion timescale is defined by
\begin{align}
    t_\mathrm{diff} := \frac{\kappa\rho}{c\lambda} \bigg(\frac{E}{|\nabla E|}\bigg)^2.
    \label{eq:tdiff}
\end{align}
In optically thick regions, this is the timescale for energy diffusion across the scale height of radiation energy density. In the optically thin limit, the flux limiter, $\lambda$, varies such that $t_\mathrm{diff}\rightarrow E/(c|\nabla E|)$, becoming an advection timescale. If $t_\mathrm{diff}$ is shorter than or comparable to the dynamical timescale of the \ac{CE} ejecta, radiation transport may alter the local pressure gradient and produce a dynamical effect\footnote{Note that $t_\mathrm{diff}$ is the timescale for the diffusion of radiation energy. In material that is dominated by gas pressure, the timescale for thermal adjustment could be significantly longer.}.

In the very beginning, only the outermost $\approx 10\%$ ($3\%$) of the \ac{RSG} envelope in mass has $t_\mathrm{diff}$ less than ten years (one year). The first column in Fig. \ref{fig:tdiff} shows that, prior to the plunge-in, low-opacity and unbound material exits through the companion's wake with $t_\mathrm{diff} \lesssim 0.01\yr$. This material is directly thrown out by the companion, acting as a gravity slingshot \citep{MacLeod+18}. Even though radiation energy diffusion is efficient in this material, it does not directly influence the energy injection process. This explains why the amount of unbound ejecta closely matches between the adiabatic and radiative simulations until the companion enters the \ac{RSG} envelope (Sect. \ref{subsec:unbound}). In Fig. \ref{fig:tdiff}b(i), although $t_\mathrm{diff}$ increases again farther from the binary, this actually indicates efficient radiation transport resulting in nearly uniform ejecta temperature. The large temperature scale length leads to a long $t_\mathrm{diff}$.

The middle column shows a snapshot taken near the end of the plunge-in. The transition between energetically bound and unbound ejecta, shown as the white region in Fig. \ref{fig:tdiff}c(ii), occurs within optically thin material. Accordingly, the evolution of the amount of unbound mass begins to diverge from that of the adiabatic case, with little to no material ejected in the radiative simulations, as noted in Sect. \ref{subsec:unbound}. The outermost tidal arms that are launched near the beginning of the dynamical plunge-in exhibit $t_\mathrm{diff}$ values less than $10^{-3}\yr$ despite still being energetically bound. This is consistent with our observation in Sect. \ref{subsec:ejecta} that these tidal arms form dense shells that are not seen in the adiabatic simulation. In contrast, most of the envelope has $t_\mathrm{diff}\sim 10\yr$, which is long relative to the inspiral timescale. The spiral shock pattern is also visible in panel b(i), due to the steep change in $E$ across a shock front.

The third column of Fig. \ref{fig:tdiff} corresponds to approximately three years after the end of the plunge-in. By this time, nearly all material in the simulation has a $t_\mathrm{diff}$ that is comparable to or shorter than the dynamical timescale, indicating a major departure from an adiabatic calculation. This is consistent with the description of the unbound plumes in Sect. \ref{subsec:ejecta}, which are smaller in extent and have denser outer shells compared to the bipolar lobes observed in the adiabatic simulation (Fig. \ref{fig:ejecta}). Radiation transport is therefore essential for accurately capturing the large-scale morphology of \ac{CE} ejecta, which could provide circumstellar medium in interacting supernovae. If such behaviour extends to lower-mass systems, incorporating radiation transport may also be necessary to correctly model the morphology of planetary nebulae.

\subsection{Location of partially ionised hydrogen}
\label{subsec:recombination}
The last row of Fig. \ref{fig:tdiff} shows the location of partially ionised hydrogen. Comparing rows 2, 4, and 5 reveals that the opacity drops sharply by a factor of $\sim 10^3$ across the hydrogen recombination front, resulting in a steep boundary beyond which $t_\mathrm{diff}$ decreases by a similar factor. The location of the hydrogen recombination front is significant because 3D adiabatic simulations of \ac{CE} evolution show that recombination energy, especially hydrogen recombination energy \citep{Lau+22b}, drastically increases the fraction of unbound material and even leads to complete envelope ejection, assuming this energy does not escape \citep[e.g.][]{Nandez+15,Sand+20,Reichardt+20,Lau+22a,Lau+22b,Vetter+24}. While our simulations do not include recombination energy release, they lend support to this key assumption since we find that the partial ionisation zone is contained within the photosphere, meaning recombination energy could be thermalised in the surrounding neutral \ac{CE} ejecta. To be converted into mechanical work, it further needs to drive expansion on a timescale that is shorter than the local diffusion timescale.

We show surfaces of constant optical depth, $\tau$, in Fig. \ref{fig:tdiff}d. These contours correspond to $\tau = 1,3,10$ in order of largest to smallest size. The optical depth is integrated along radial rays from infinity to the origin using the same opacity calculated for radiative diffusion. In Fig. \ref{fig:tdiff}e, the contours show optical depths that only include the contribution from bound material, $\tau_\mathrm{bound}$, to assess the ability for recombination radiation to assist with ejection. The $\tau_\mathrm{bound}$ contours are larger and have different shapes compared to the $\tau$ contours in Fig. \ref{fig:tdiff}d, implying that the unbound ejecta at larger radii have non-negligible optical thicknesses and exhibit asymmetry\footnote{We have not accounted for dust formation by the material previously ejected through the $L_2$ point, which likely forms an obscuring photosphere at larger radii. In their 3D \ac{CE} simulation that self-consistently models dust nucleation and growth, \cite{Bermudez+24} find that dust is mainly formed in unbound envelope ejecta, implying that any recombination energy absorbed by these layers would not help with \ac{CE} ejection.}. In Fig. \ref{fig:tdiff}e~(ii) \& (iii), the ionisation zone is entirely contained within the $\tau_\mathrm{bound} = 10$ contour. The result that hydrogen recombines inside the photosphere agrees with Paper I and has also been reported in other studies \citep[e.g.][]{Sand+20,Bronner+24}. It is important to verify whether the photosphere still contains the hydrogen partial ionisation zone when incorporating both radiation transport and recombination energy release. Recombination energy absorbed in optically thick regions drives additional expansion, which could cause the material to become optically thin again. Our previous adiabatic simulations that include recombination energy show that the hydrogen partial ionisation zone is much wider, of the order of thousands of solar radii, after the plunge-in (see Fig. 13 of Paper I). Simulations incorporating radiation transport and recombination energy release will also reveal the efficiency with which thermalised energy can be converted to mechanical work.

%% file: discussion.tex
\section{Discussion} \label{sec:discussion}

\subsection{Limitations}
\label{subsec:limitations}
There are important limitations to our implementation of energy transport, particularly the neglect of pre-existing envelope convection, the lack of explicit cooling at the photosphere, and shortcomings of the flux-limited diffusion approximation. Pre-existing envelope convection was deliberately suppressed through adopting a flat entropy profile in the \ac{RSG} envelope (Sect. \ref{subsec:thermal_structure}). This choice reflects the assumption that such convection transports the \ac{RSG}'s nuclear luminosity, which is not included in the default `radiative' simulation. At first order, the energy injected during the spiral-in must be transported by additional convective flows. Pre-existing convection may be important if it interferes non-linearly with these flows, and if they are not disrupted by spiral shocks that develop during the plunge-in. Several studies have argued that additional transonic or supersonic convection could transport away orbital energy and hinder envelope ejection \citep[e.g.][]{Sabach+17,Wilson&Nordhaus20,Wilson&Nordhaus22,Noughani+24}. However, many of these works rely on simplifying assumptions such as mixing-length theory, spherical symmetry, and static stellar models. Their validity should be verified with self-consistent 3D simulations.

Additionally, the diffusion approximation could lead to unphysical effects when applied to an optically thin medium. The flux limiter ensures that the radiative flux approaches the correct asymptotic limit, $|\mathbf{F}_\mathrm{rad}| \rightarrow cE$, but radiation energy can nonetheless only be transported anti-parallel to the temperature gradient, prohibiting the formation of shadows behind opaque obstacles. We also neglected photon scattering, assumed \ac{LTE}, and adopted gray opacities. More sophisticated methods such as Monte Carlo radiative transfer \citep{Whitney11}, variable Eddington tensor schemes \citep{Stone+92}, and discrete ordinates methods \citep[e.g.][]{Jiang21,Ma+25} mitigate some of these problems, but are significantly more computationally expensive and challenging to implement. It would be valuable to verify our preliminary results with these approaches in the future.

Another challenge is accurately capturing cooling near the start of the simulation, which has two main difficulties. Firstly, resolving the steep temperature stratification at the \ac{RSG} photosphere requires prohibitively high resolution. This long-standing issue is known to severely overestimate the effective temperature and luminosity in post-processed light curves from 3D \ac{CE} simulations (\citealt{Galaviz+17}, but, see \citealt{Hatfull+21} and \citealt{Hatfull+24}). Secondly, the surface layer of \ac{SPH} particles cannot radiate into the background `vacuum' because the radiative diffusion method only allows energy exchange between particles. Modelling photospheric cooling is necessary for driving pre-existing convection in the \ac{RSG} envelope. Potential solutions include implementing a separate treatment for surface cooling or introducing boundary particles to absorb energy radiated by the surface layer. However, these issues mainly affect the pre-plunge-in evolution in a \ac{CE} simulation, as \ac{CE} ejecta eventually provides material into which the photosphere can cool.

A final caveat is that recombination energy has not been modelled together with radiation transport. As discussed in Sect. \ref{subsec:recombination}, adiabatic simulations show that recombination energy enhances envelope ejection and increases the post-plunge-in orbital separation. In Paper II, we demonstrated that helium and hydrogen recombination energy increase the ejected mass by 48\% and the final separation by 16\%. Hydrogen recombination energy also leads to forming more extended and spherically symmetric ejecta. It is therefore important to check whether radiation transport interferes with recombination-assisted envelope ejection.

\subsection{Comparison with related works}
\label{subsec:related_works}
At the time of writing, few studies have incorporated radiative diffusion into detailed \ac{CE} simulations and systematically analysed its effects. In a conference proceedings contribution, \cite{Ricker+18} presented preliminary results from a \ac{CE} simulation involving an 82.1\Msun \ac{RSG} donor and a 35\Msun black hole companion. This simulation, conducted using the 3D adaptive mesh refinement code {\scshape flash} with flux-limited diffusion, showed that vigorous envelope convection observed in an adiabatic setup is suppressed when radiative diffusion is included. They attribute this suppression to radiation carrying part of the energy flux that would otherwise be transported entirely by convection. In our setup, envelope convection does not develop before the plunge-in, and so we are unable to make the same comparison. However, as discussed in Sect. \ref{subsec:ejecta}, we observe fewer small-scale flows in the turbulent ejecta after the plunge-in phase in the radiative simulation compared to the adiabatic case.

Recently, \cite{Hatfull+24} implemented a `flux-limited emission-diffusion' technique into the \ac{SPH} code \texttt{StarSmasher}. This was applied to simulating the merger of V1309 Sco in 3D and computing its light curve. Unlike standard flux-limited diffusion, their method does not include radiation momentum in the momentum equation (Eq. (\ref{eq:euler})). However, this term is not dynamically relevant for the typical timescales and luminosities associated with stellar mergers and dynamical \ac{CE} evolution \citep{Glanz&Perets18,Bermudez+24}. Their approach to radiative diffusion also differs from the standard \ac{SPH} discretisation used by \cite{Whitehouse+Bate04} and \cite{Whitehouse+05}, instead relying on energy exchange between neighbouring particles that overlap with discrete rays traced from each particle. This allows cooling into empty regions devoid of particle neighbours. Although this is not possible with our method, as noted in Sect. \ref{subsec:limitations}, the photosphere in our simulations can still cool to surrounding optically thin material once the \ac{CE} expands significantly. Before such expansion, the steep temperature gradient at the \ac{RSG} photosphere is unresolved in any case, preventing calculation of the correct cooling rate.

\cite{Pejcha+16b,Pejcha+16a} conducted 3D \ac{SPH} simulations of $L_2$ mass outflows that incorporated radiative diffusion and a prescription for radiative cooling. They identified cases where bound gas falls back towards the binary and undergoes shock heating. In the limit of efficient cooling, this gas forms a circumbinary disk and a fraction of the heated gas escapes along the unobstructed poles. Despite our dissimilar setups, these findings are qualitatively similar to the bound \ac{CE} ejecta and poloidal outflows discussed in Sect. \ref{subsec:ejecta}.

A number of 1D \ac{CE} simulations have included energy transport \citep{Meyer+Meyer-Hofmeister79,Fragos+19,Bronner+24}, though they often do not isolate the effects of radiative diffusion by directly comparing with an adiabatic simulation. \cite{Fragos+19} performed 1D \ac{CE} hydrodynamic simulations involving a 12\Msun \ac{RSG} donor and a 1.4\Msun neutron star companion. They incorporated radiative diffusion and convective energy transport using mixing length theory. Their model also included various energy sources and sinks, most notably, radiative losses from the photosphere, nuclear energy, and recombination energy. They found that the system transitioned to self-regulated evolution several years after the end of the plunge-in.

Both \cite{Fragos+19} and \cite{Bronner+24} showed that the majority of the envelope mass may be ejected, which starkly contrasts with the present work. However, the outcomes of 1D simulations are sensitive to assumptions regarding the size of the energy deposition region and the manner in which energy is deposited into the 1D stellar model. For instance, one extreme case in \cite{Bronner+24} required depositing energy in a shell four times thicker than the companion's Bondi-Hoyle-Lyttleton sphere of influence to reproduce the orbital evolution seen in a 3D simulation. The inherent instantaneous energy homogenisation over spherical shells in 1D models may also overestimate mass ejection \citep[though, see a recent approach to address this by][]{Everson+25}. Taken together, these uncertainties could plausibly explain the differences between 1D and 3D simulation results.

\subsection{Implications for binary stellar evolution}
\label{subsec:implications}
Today, \ac{CE} evolution remains a major bottleneck in understanding the formation of compact binaries almost half a century since its initial proposal \citep{Paczynski1976}. Particularly, it is questionable whether 3D simulations have self-consistently demonstrated \ac{CE} ejection. Simulations that do not model recombination energy release typically find that most of the envelope mass (70--90\%) remains bound upon termination of the dynamical plunge-in \citep[e.g.][]{Lau+22a}. In contrast, simulations that include recombination energy show that most if not all of the envelope could be fully ejected over many years after the plunge-in \citep[e.g.][]{Nandez+15,Sand+20,Reichardt+20,Lau+22a,Lau+22b,Vetter+24}. However, these simulations are all adiabatic, and whether complete ejection could be demonstrated with radiation transport remains an open question. One key debate centres on whether hydrogen recombination energy contributes to envelope ejection or is mostly radiated away \citep[e.g.][]{Sabach+17,Grichener+18,Ivanova18,Soker+18}. Our results indicate that non-adiabatic effects influence both envelope ejection and ejecta morphology shortly after the plunge-in. The still-bound portion of the \ac{CE} may be transitioning towards self-regulation \citep{Meyer+Meyer-Hofmeister79,Podsiadlowski01} and it is unknown whether this still allows recombination-driven ejection, such as the radial disk winds characterised by \cite{Vetter+24}.

On the other hand, 3D adiabatic simulations generally demonstrate that an immediate merger of the stellar cores may be averted even if most of the envelope remains bound. The bound material instead forms an extended, centrifugally supported structure around the central binary \citep{Kuruwita+16,Gagnier&Pejcha23,Tuna+Metzger23,Vetter+24}. Embedded in low-density, co-rotating gas, the central binary experiences slow inspiral, with an orbital decay timescale that is a thousand times longer than the orbital period \citep{Lau+22a}. Our radiative simulations show even more bound material persisting after the spiral-in compared to the adiabatic case, suggesting an even more massive disk. These results further challenge the prevalent picture of \ac{CE} evolution involving a dynamical-timescale phase that rapidly shrinks the orbital separation while ejecting the entire envelope. Instead, most of the envelope is likely removed through a long-timescale, non-adiabatic process that may further influence the central binary. Observations of post-\ac{CE} systems with distant tertiary companions support mass-loss timescales of $10^2-10^4\yr$ \citep{Michaely+Perets19,Igoshev+20}. This also suggests that the widely used `energy formalism' or `$\alpha$-formalism' should not be applied to the binding energy of the entire hydrogen-rich envelope.

Alternative models for envelope ejection have been proposed. Recently, \cite{Hirai+Mandel22} suggested dividing the envelope ejection process into an adiabatic and non-adiabatic phase, with the convective-radiative boundary in the envelope serving as a possible division point. Their formalism translates into a \ac{CE} efficiency parameter that is strongly dependent on system properties, ranging from $\alpha \sim 0.1$ to $>10^4$. Another line of research explores how the post-\ac{CE} binary orbit evolves under the influence of a leftover circumbinary disk \cite[e.g.][]{Kashi+Soker11,Tuna+Metzger23,Izzard+Jermyn23,Wei+24,Valli+24,Unger+24}. If a substantial fraction of the donor star's envelope ($\gtrsim 10\%$ of the post-\ac{CE} binary mass) forms a circumbinary disk, its angular momentum could significantly alter the post-\ac{CE} orbital separation and induce mild orbital eccentricities \citep[e.g.][]{DOrazio+Duffell21,Siwek+23a,Siwek+23b}. This has far-reaching implications for binary evolution. For example, in gravitational wave sources like binary black holes and double neutron stars, circumbinary disk torques may be necessary in certain cases to sufficiently shrink the post-\ac{CE} orbit, allowing a merger to take place within a Hubble time \citep{Moreno+22,Wei+24,Vetter+24,Landri+25}. Beyond the \ac{CE} phase, circumbinary disks could also drive a subsequent case BB mass transfer episode to become unstable and provide circumstellar medium in interacting supernovae \citep{Tuna+Metzger23,Ercolino+24,Wei+24}. For example, SN 2014c is consistent with interaction with toroidal circumstellar material that may have been ejected through \ac{CE} evolution \citep{Brethauer+22,Orlando+24}.

%% file: conclusion.tex
\section{Summary and conclusions} \label{sec:conclusion}
We performed 3D radiation hydrodynamics simulations of the \ac{CE} phase involving a 12\Msun \ac{RSG} and a 3\Msun companion, using an implicit two-temperature flux-limited diffusion solver implemented in the \ac{SPH} code \Phantom. By comparing these simulations to our previous adiabatic simulations, we assessed the impact of radiation transport on envelope ejection, ejecta morphology, and the final orbital separation. Our key conclusions are listed below:

\begin{enumerate}[(i)]
    \item Radiative diffusion significantly obstructs envelope ejection, reducing the unbound mass to less than half of that in the adiabatic simulation.
    \item The immediate post-plunge-in orbital separation is largely unaffected by radiative diffusion.
    \item The envelope mass ejected in the companion's wake before the plunge-in is not affected by radiative diffusion. However, radiative diffusion causes little to no mass to be ejected during the plunge-in itself. Late-time ejection is also suppressed due to ejecta falling back towards the central binary, obstructing further mass outflow. During these phases, the outermost bound ejecta have short radiation transport timescales due to the opacity drop associated with hydrogen recombination.
    \item While poloidal outflows inflate clear bipolar lobes in the adiabatic simulation, the radiative case instead produces intermittent and sometimes single-sided plumes. As a result, the bipolar morphology is either absent or significantly less pronounced.
    \item Although our simulations do not include recombination energy release. The hydrogen (and helium) partial ionisation zones have optical depths exceeding ten. This supports the notion that recombination radiation could thermalise in bound ejecta. Whether it can contribute to the expansion of the \ac{CE} remains an open question that should be addressed in future simulations incorporating both recombination energy release and radiation transport.
\end{enumerate}

This study represents a preliminary step towards understanding the role of radiation transport in \ac{CE} evolution. As discussed in Sect. \ref{subsec:limitations}, the main limitations of our simulations include the breakdown of the flux-limited diffusion approximation in optically thin regions, difficulty capturing photospheric cooling near the start of the simulation, and the omission of recombination energy release. Addressing these limitations would be a crucial next step. It would also be valuable to validate our results using different hydrodynamics codes and radiation transport methods. Since our findings are based on a specific binary system, it would be instructive to explore the impact of radiative diffusion with different mass ratios and evolutionary stages of the donor star. Going forward towards the paradigm of partial dynamical \ac{CE} ejection, the orbital separation and bound envelope mass recorded at the end of 3D simulations can serve as initial conditions for dedicated simulations of the subsequent non-adiabatic evolution.

%% file: resolution.tex
\section{Energy and angular momentum conservation}
\label{app:conservation}
\begin{figure}
    \centering
    \includegraphics[width=\linewidth]{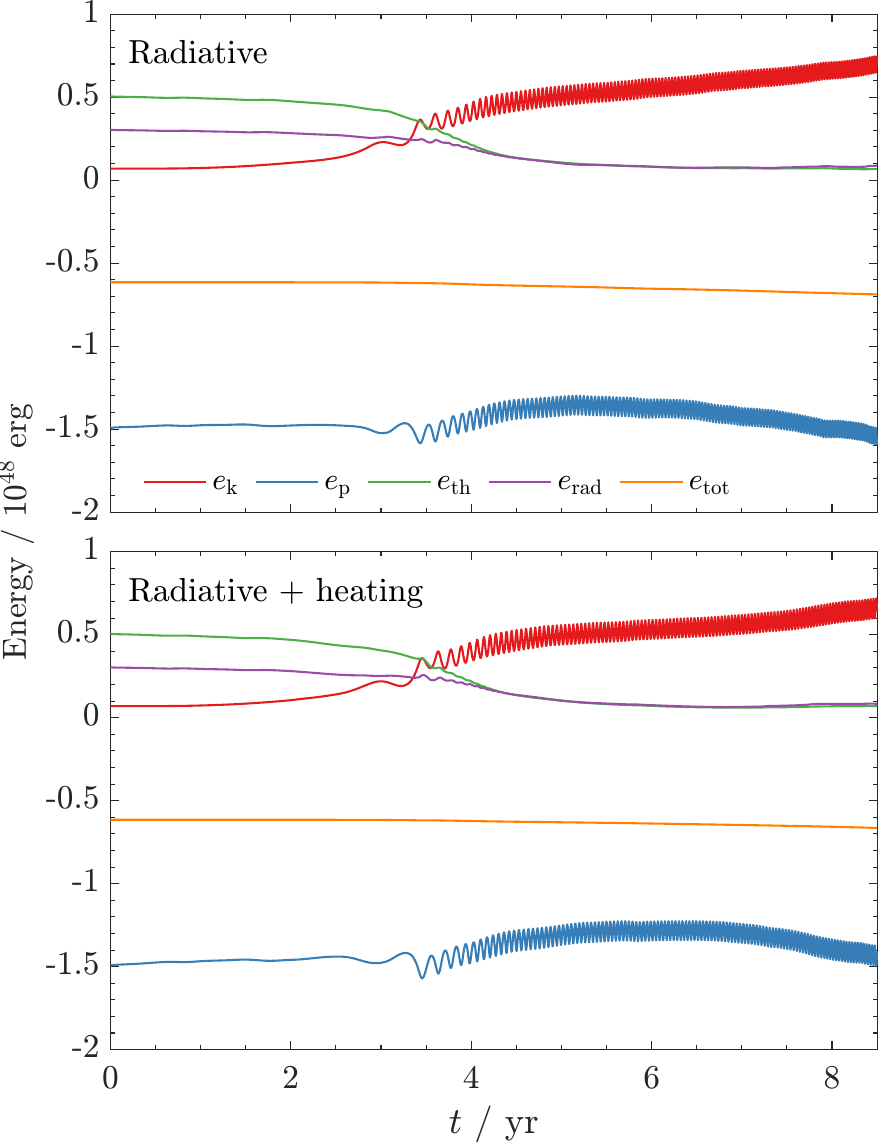}
    \caption{Time evolution of kinetic energy (red), gravitational potential energy (blue), gas thermal energy (green), radiation energy (purple), and total energy (orange). Top panel: `Radiative' simulation. Bottom panel: `Radiative + heating' simulation.}
    \label{fig:energy_conservation}
\end{figure}

Total energy is conserved within 10.7\% for the `radiative' simulation and 6.9\% for the `radiative + heating' simulation up to $t=8\yr$. In the latter case, this does not include energy explicitly added to model the \ac{RSG}'s nuclear luminosity, which totals $4.2\times10^{46}\erg$ over 8\yr. The evolution of different energy components is shown in Fig. \ref{fig:energy_conservation}, with most of the total energy drift occurring after the plunge-in. Energy conservation is significantly worse than in the adiabatic simulation from Paper I, which maintained total energy to within 0.05\%. This difference arises from the implicit radiation solver, which ensures numerical stability with arbitrarily large time steps but does not conserve energy exactly. The relatively strong drift in total energy after the plunge-in could be associated with radiative diffusion in optically thin ejecta. Lowering the accuracy tolerance of the implicit solver could improve energy conservation. In contrast, angular momentum is conserved within 0.020\% for the `radiative' simulation and 0.022\% for the `radiative + heating' simulation, comparable to the adiabatic case.

\section{Resolution test}
\begin{figure}
    \centering
    \includegraphics[width=\linewidth]{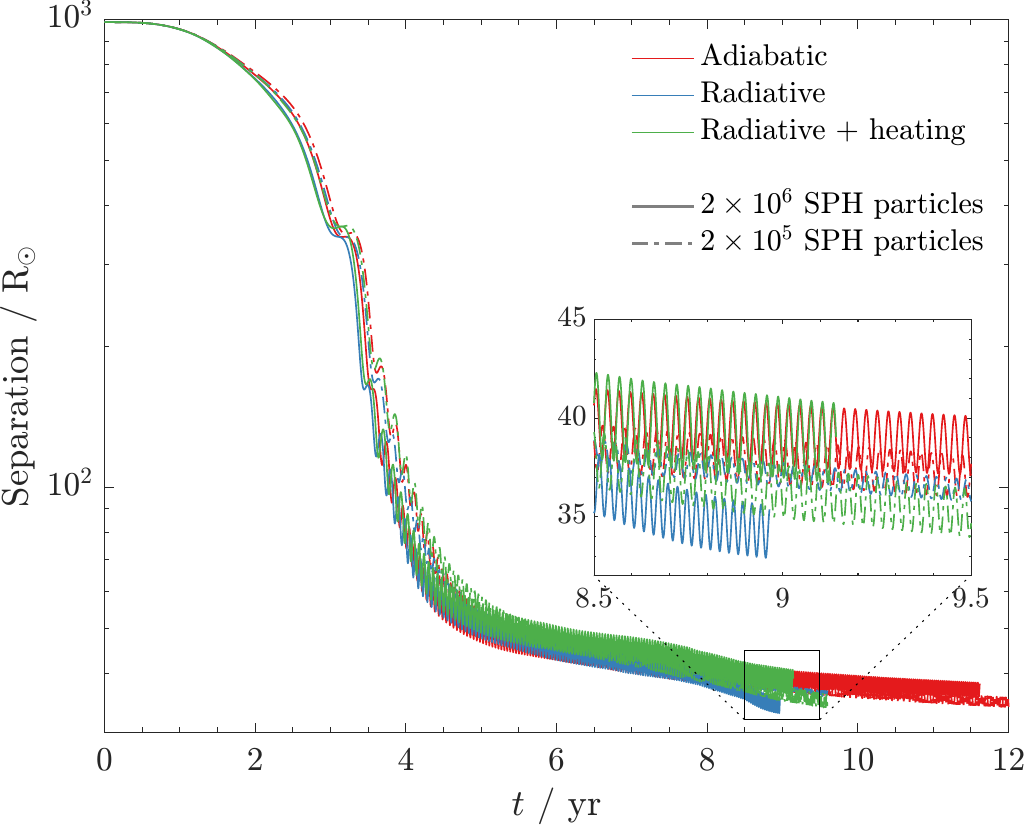}
    \caption{Separation of the stellar cores as a function of time, similar to Fig. \ref{fig:sep}, but including results obtained with ten times fewer \ac{SPH} particles (dash-dotted lines).}
    \label{fig:sep_res}
\end{figure}

\begin{figure}
    \centering
    \includegraphics[width=\linewidth]{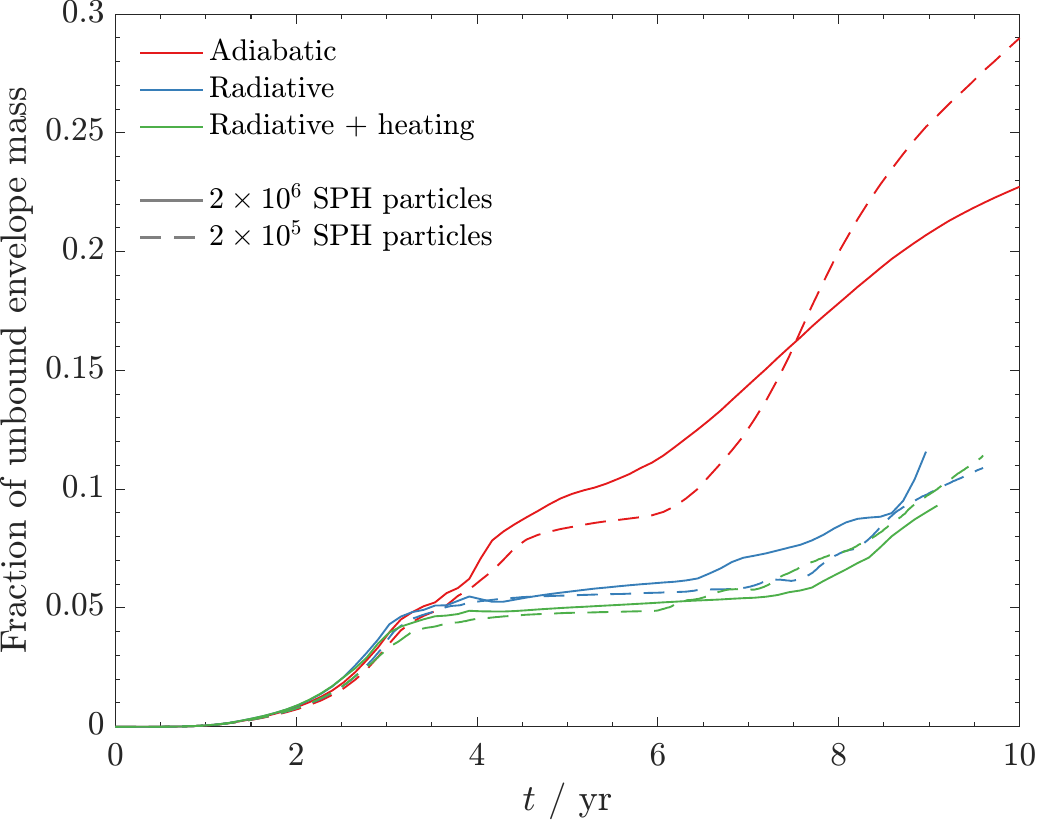}
    \caption{Amount of unbound mass as a function of time, similar to Fig. \ref{fig:unbound}, but including results obtained with ten times fewer \ac{SPH} particles (dashed lines). Material is considered to be unbound if the sum of its kinetic and potential energy is positive ($e_\mathrm{k} + e_\mathrm{p} > 0$).}
    \label{fig:unbound_res}
\end{figure}

To assess the sensitivity of our results to resolution, we compare the evolution of orbital separation (Fig. \ref{fig:sep_res}) and unbound mass (Fig. \ref{fig:unbound_res}) using a lower-resolution run with ten times fewer \ac{SPH} particles ($N = 2\times10^5$). For the purely adiabatic case, we display results from \cite{Lau+22b}, corresponding to the simulation labelled `None, fixed $\mu$'.

In Fig. \ref{fig:sep_res}, all simulations show similar orbital separation evolution until the pronounced drop in the `radiative' case near 8.5\yr, which was discussed in Sect. \ref{subsec:final_sep}. This drop in the orbital separation is absent in the lower-resolution run, suggesting it may be a stochastic effect caused by turbulent ejecta in the slow spiral-in or a numerical artefact. Despite this, there is little resolution dependence in the orbital separation evolution for at least $\approx1-2$ years after the plunge-in, preserving the key result that radiative diffusion has minimal impact on the post-plunge-in separation. At lower resolution, the onset of the plunge-in is delayed by $\approx 0.1\yr$ in all cases. This may be due to a small amount of unresolved surface mass on the \ac{RSG}, which delays the onset of mass transfer that initially contracts the orbit.

Figure \ref{fig:unbound_res} compares the evolution of unbound mass across the two resolutions. In the `radiative' case, the results agree within 7\% in the first six years. Larger deviations arise during intermittent plume ejections (Sect. \ref{subsec:ejecta}), when spatial resolution in the binary's vicinity decreases due to lower density. The adiabatic case also shows good agreement until $\approx 4\yr$, around the transition towards a slower inspiral. At both resolutions, the adiabatic simulation exhibits sustained mass ejection after the plunge-in that is much weaker with radiative diffusion, suggesting that this qualitative difference is robust.

%% file: test_problems.tex
\section{Test problems for radiation hydrodynamics}
\label{app:tests}
We present results of selected radiation hydrodynamics test problems following \cite{Whitehouse+Bate04} and \cite{Whitehouse+05}. An important distinction is that our tests are performed in 3D with \ac{SPH} particles arranged in a hexagonal close-packed lattice. For tests involving a discontinuity (Appendices \ref{app:radpulse} and \ref{app:radshock}), the setup and boundary conditions follow the description in \cite{Price+18} for the shock tube test. In these tests, the discontinuity is defined on a constant-$x$ plane, with periodic boundaries imposed in the $y$- and $z$-directions. The first and last layers of \ac{SPH} particles along the $x$-direction are tagged as boundary particles so that their properties are fixed. The results presented here can be reproduced using the \Phantom test suite and are subject to regular unit testing. In all tests, we assume an ideal gas with an adiabatic index of $\gamma = 5/3$ and a mean molecular weight of $\mu=2$. We set an accuracy tolerance of $10^{-6}$ for the implicit solver.

\subsection{Gas-radiation energy exchange}
We test the gas and radiation energy exchange terms in Eqs. (\ref{eq:rad}) and (\ref{eq:u}). In the first test, we initialise a static, uniform-density, and optically thick gas with $T_\mathrm{gas} \gg T_\mathrm{rad}$, allowing radiative emission from the gas to heat the radiation until thermodynamic equilibrium is reached ($T_\mathrm{gas} = T_\mathrm{rad}$). The gas is assumed to have a fixed opacity of $\kappa=0.4\cmsg$. The initial conditions are $u = 10^{10}\ergg$, $\rho = 10^{-7}\gcc$, and $\xi=10^{12}\ergg$. Unlike other tests, \ac{SPH} particles are arranged in a cubic lattice within in a cube of side length $1\AU$ with periodic boundaries.

\begin{figure}
    \centering
    \includegraphics[width=\linewidth]{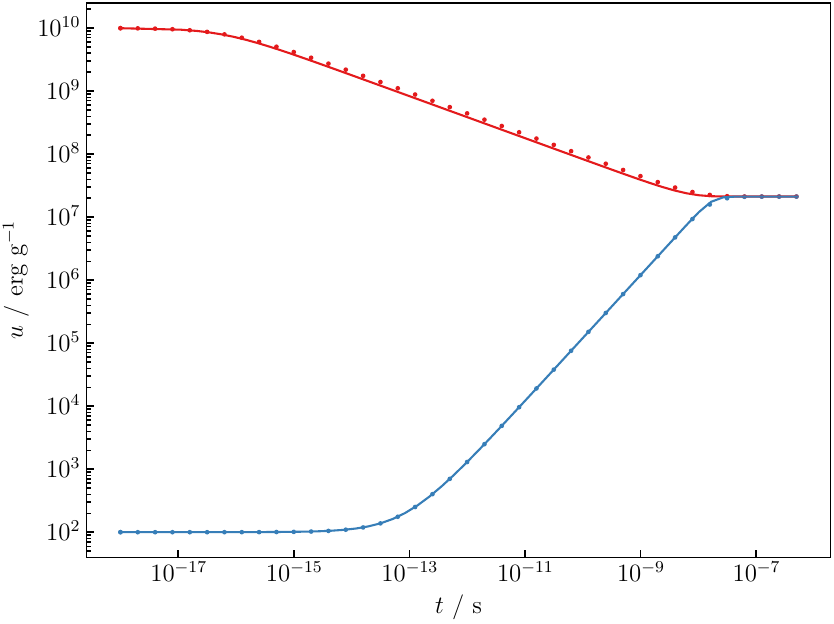}
    \caption{Evolution of thermal energy in the tests for the gas-radiation exchange terms. Markers show results obtained with the implicit radiation solver with forty logarithmically spaced time steps, which are in close agreement with results obtained from an explicit solver (solid lines). The case of gas cooling ($T_\mathrm{gas}\gg T_\mathrm{rad}$) is plotted in red while the case of gas heating ($T_\mathrm{gas} \ll T_\mathrm{rad}$) is plotted in blue.}
    \label{fig:gas_exchange}
\end{figure}

A second test is set up identically, but with $T_\mathrm{gas} \ll T_\mathrm{rad}$, so that the gas is instead heated towards the radiation temperature. In this case, the initial gas thermal energy is set to be $u=10^2\ergg$. Figure \ref{fig:gas_exchange} shows the evolution of thermal energy for both tests. The results obtained with the implicit solver using 40 time steps (plotted as markers) closely match those obtained with an explicit solver (solid lines).

\subsection{Diffusion of a sinusoid}
\begin{figure}
    \centering
    \includegraphics[width=\linewidth]{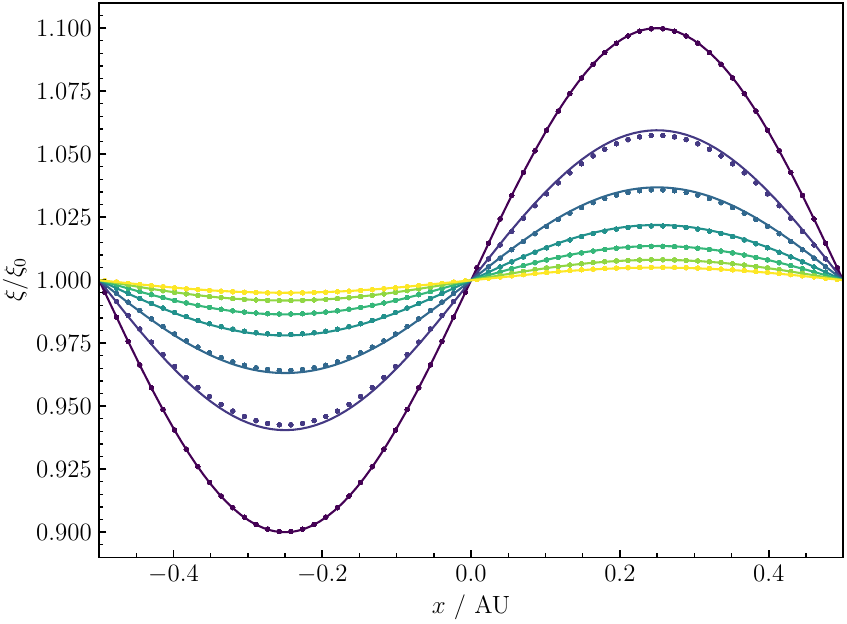}
    \caption{Diffusion of sinusoidally varying specific radiation energy. The markers represent results obtained from the implicit radiation solver, while solid lines show the analytical solution to the heat equation. The solutions at seven equally spaced times are plotted, with darker lines indicating earlier times and lighter lines representing later times.}
    \label{fig:sinusoid_diffusion}
\end{figure}

The radiative diffusion term in Eq. (\ref{eq:rad}) is tested by simulating diffusion in a periodic particle lattice with a sinusoidally varying $\xi$. This test was implemented in \Phantom by \cite{Biriukov2020}. We disable the gas-radiation exchange terms tested previously and set the flux limiter to the optically thick value, $\lambda=1/3$. A fixed time step of $\Delta t = 3.37\times 10^{-17}\seconds$ is used. Particles are placed in a uniform lattice of size $x \in [-0.5,0.5]~\AU$, $y,z\in [-0.1,0.1]~\AU$, and mass density of $\rho = 1.49\times10^{-30}\gcc$. The opacity is fixed at the value $\kappa = 1\cmsg$. The initial specific radiation energy varies sinusoidally along the $x$-direction as $\xi(t=0) = \xi_0 [1+0.1\sin(2\pi x)]$, where $\xi_0 = a_\mathrm{rad} T_\mathrm{rad,0}^4/\rho$ with $T_\mathrm{rad,0}=100~\mathrm{K}$. As there is no fluid motion, Eq. (\ref{eq:rad}) reduces to a heat equation, for which the analytical solution follows an exponential decay towards $\xi_0$, $\xi(t) = \xi_0 [1+0.1\sin(2\pi x)\exp(-4\pi^2 D t)]$, where $D=c/(3\kappa\rho)$ is the diffusion coefficient. Figure \ref{fig:sinusoid_diffusion} shows close agreement between the analytical solution and the results obtained from the implicit radiation solver, with a maximum relative deviation of 0.2\%.

\subsection{Radiation pulse in an optically thin medium}
\label{app:radpulse}
\begin{figure}
    \centering
    \includegraphics[width=\linewidth]{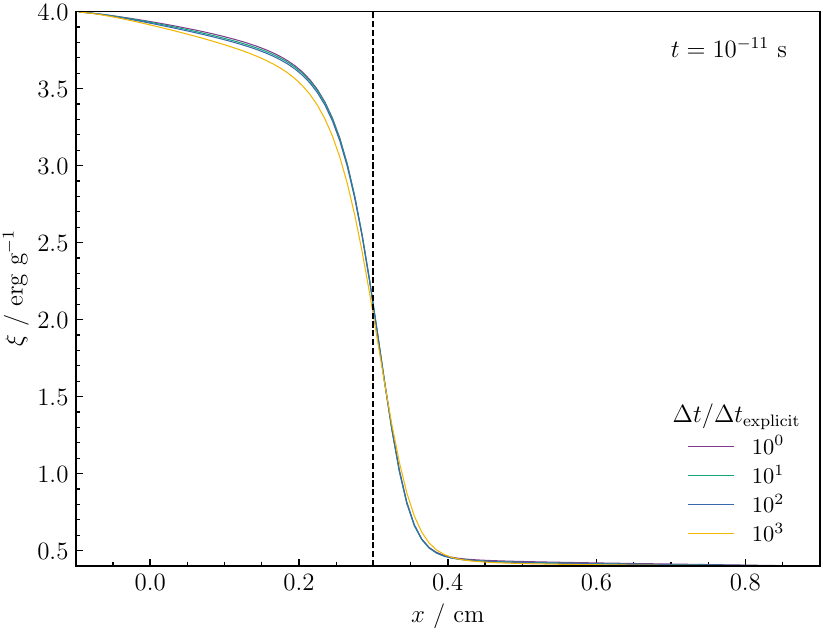}
    \caption{Propagation of a radiation pulse in an optically thin medium computed with different time steps ranging from the explicit time step to one thousand times the explicit time step. The discontinuity in $\xi$ propagates from $x=0$ at $t=0$ to the expected position of $ct=0.3\cm$ (dotted line) at $t=10^{-11}\seconds$.}
    \label{fig:radpulse}
\end{figure}

We simulate a radiation pulse propagating in optically thin material to verify that the radiation flux is correctly limited by the speed of light. The setup consists of gas and radiation initially in thermal equilibrium in a box with dimensions $x\in [-0.1, 0.9]\cm$ and $y,z\in[-0.1,0.1]\cm$. The gas density is 0.025\gcc and the opacity is fixed at 0.4\cmsg. The radiation energy is initially $\xi = 0.4 \ergg$. At $t=0$, radiation energy in the region $x<0$ is instantaneously increased by a factor of ten to $\xi = 4\ergg$. Layers of boundary particles are then used to fix $\xi$ at the $x$-boundaries. Figure \ref{fig:radpulse} shows the $\xi$-profile along the $x$-direction at $t=10^{-11}\seconds$, along with the expected pulse position of $ct=0.3\cm$ (dotted line). Results are displayed for implicit time steps $\Delta t$ equal to 1, 10, 100, and 1000 times the explicit time step, $\Delta t_\mathrm{explicit}$. The $\xi$-profiles closely agree, with maximum errors relative to the case of $\Delta t/\Delta t_\mathrm{explicit} = 1$ of 1.3\%, 1.6\%, and 4.3\% for $\Delta t/\Delta t_\mathrm{explicit} = 10$, 100, and 1000, respectively.

\subsection{Adiabatic and isothermal shocks}
\label{app:radshock}
\begin{figure*}
    \centering
    \includegraphics[width=\linewidth]{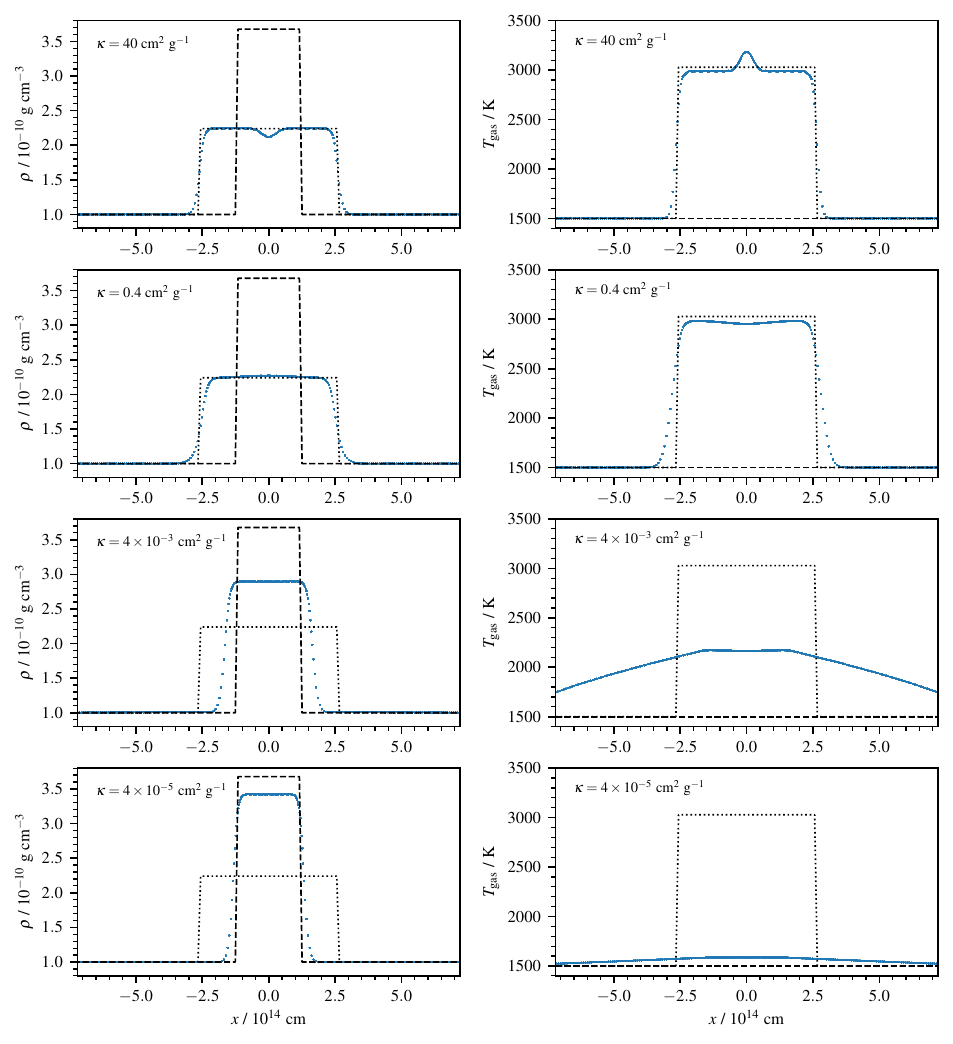}
    \caption{Solutions for the radiative shock at $t=10^9~\mathrm{s}$ for different opacities. The top row corresponds to the optically thick (adiabatic) case, gradually transitioning to the optically thin (isothermal) case in the bottom row. The dotted line shows the analytical solution for the adiabatic case while the dashed line corresponds to the isothermal solution. The left column shows density profiles, and the right column shows gas temperature profiles.}
    \label{fig:radshock}
\end{figure*}

We set up a shock tube with varying opacities to explore different shock regimes, ranging from adiabatic to nearly isothermal. Similar to Appendix \ref{app:radpulse}, we place particles in an elongated 3D box with a length of $x\in[-10^{15},10^{15}]\cm$. The gas has a density of $\rho_0 = 10^{-10}\gcc$ and is initially in thermal equilibrium with a radiation field at the temperature $T_0=1500\K$. Material in the region $x<0$ is given an initial velocity of $v_0=3.2\times 10^5\cms$, while material in $x>0$ is given a velocity of $-v_0$. The flows collide at $x=0$ and form a shock. We simulate the shock tube with varying gas opacities, $\kappa = 40,0.4,4\times10^{-3},4\times10^{-5}\cmsg$, progressively transitioning from an adiabatic to a nearly isothermal shock. Figure \ref{fig:radshock} shows the density and temperature distribution at $t=10^{9}\seconds$ for each opacity. The analytical solution for the adiabatic (dotted line) and isothermal (dashed line) shocks are also shown for comparison. The shock speed may be obtained from the Rankine-Hugoniot jump conditions as 
\begin{align}
    D = \frac{1}{4} \bigg[ (\gamma_\mathrm{eff}-3) + \sqrt{\frac{16\gamma_\mathrm{eff} p_0}{\rho_0v_0^2} + (\gamma_\mathrm{eff}+1)^2} \bigg]v_0,
\end{align}
where $p_0$ is the pre-shock pressure and $\gamma_\mathrm{eff}$ is the effective adiabatic index, assuming the value of $5/3$ in the adiabatic limit and $1$ in the isothermal limit. The post-shock density, $\rho_1$, is given by
\begin{align}
    \frac{\rho_1}{\rho_0} = 1-\frac{v_0}{D},
\end{align}
while the post-shock temperature in the optically thick case is given by
\begin{align}
    \frac{T_1}{T_0} = \frac{\rho_0}{\rho_1} - \frac{\rho_0 v_0 D}{p_0}
\end{align}
in the limit where gas pressure dominates. Figure \ref{fig:radshock} closely reproduces the results of Fig. 3 from \cite{Whitehouse+05} and demonstrates agreement with both the adiabatic and isothermal solutions. In the top panels, the temperature bump and associated density dip near $x=0$ are due to wall heating \citep{Noh87,Rider00} and are also seen in \cite{Whitehouse+05}. They are smoothed out in the other panels due to more efficient radiative diffusion when approaching the isothermal limit.

%% file: movies.tex
\section{Movies of simulations}
\label{app:movies}

\makegapedcells
\begin{table}[h!]
\caption{Simulation videos with compressed versions accessible via \protect\url{https://themikelau.github.io/radiation_CE.html} and full-resolution versions available via \protect\url{https://zenodo.org/uploads/15544695}. }
    {
    \centering
    \begin{tabular}{@{}ll@{}}
        \toprule
        Simulation(s)           & View \\ \midrule
        Adiabatic            & \makecell{Equatorial \\ Meridional \\ Co-rotating} \\
        Radiative            & \makecell{Equatorial \\ Meridional \\ Co-rotating} \\
        Radiative + heating  & \makecell{Equatorial \\ Meridional \\ Co-rotating} \\
        Side-by-side comparison & \makecell{Equatorial \\ Meridional} \\
        \makecell{Side-by-side comparison \\ (adiabatic \& radiative only)} &  \makecell{Equatorial \\ Meridional} \\
        \bottomrule
    \end{tabular}
    \par
    }
    \label{tab:movies}
    \tablefoot{Each movie shows the time evolution of a density cross-section. The equatorial view is in the $xy$-plane (the initial orbital plane), while the meridional view is in the $xz$-plane. The co-rotating view is in a rotating meridional plane that always contains both stellar cores. Movies for the adiabatic simulation are same as those for the gas + radiation \ac{EoS} simulation in Paper I \citep{Lau+22a}.}
\end{table}